# High-pressure structural study of the scheelite tungstates $CaWO_4$ and $SrWO_4$

D. Errandonea,[1*] J. Pellicer-Porres,[1] F. J. Manjón,[2] A. Segura,[1] Ch. Ferrer-Roca,[1]

R. S. Kumar,[3] O. Tschauner,[3] P. Rodríguez-Hernández,[4] J. López-Solano,[4] S. Radescu,[4]

A. Mujica,[4] A. Muñoz,[4] and G. Aquilanti[5]

[1] Departamento de Física Aplicada-ICMUV, Universitat de València,
Edificio de Investigación, c/Dr. Moliner 50, 46100 Burjassot (Valencia), Spain

[2] Departamento de Física Aplicada, Universitat Politècnica de València,
Cno. de Vera s/n, 46022 València, Spain

[3] High Pressure Science and Engineering Center, Department of Physics,
University of Nevada, 4505 Maryland Parkway, Las Vegas, Nevada 89154-4002, USA

[4] Departamento de Física Fundamental II, Universidad de La Laguna, La Laguna 38205, Tenerife, Spain

[5] European Synchrotron Radiation Facility, BP 220, Grenoble, F-38043 France

**Abstract:** Angle-dispersive x-ray diffraction (ADXRD) and x-ray absorption near-edge structure (XANES) measurements have been performed on $CaWO_4$ and $SrWO_4$ up to pressures of approximately 20 GPa. Both materials display similar behavior in the range of pressures investigated in our experiments. As in the previously reported case of $CaWO_4$, under hydrostatic conditions $SrWO_4$ undergoes a pressure-induced scheelite-to-fergusonite transition around 10 GPa. Our experimental results are compared to those found in the literature and are further supported by *ab initio* total energy calculations, from which we also predict the instability at larger pressures of the fergusonite phases against an orthorhombic structure with space group *Cmca*. Finally, a linear relationship between the charge density in the $AO_8$ polyhedra of $ABO_4$ scheelite-related structures and their bulk modulus is discussed and used to predict the bulk modulus of other materials, like hafnon.

* Corresponding author; electronic mail: daniel.errandonea@uv.es, Tel.: (34) 96 354 34 32, FAX: (34) 354 3146

## I. Introduction

Scheelite $ABX_4$ compounds are important materials from both a theoretical and a technological point of view. Scheelite fluorides ($ABF_4$) like $YLiF_4$ and $GdLiF_4$ are used in rare-earth solid state lasers **[1]**, scheelite oxides ($ABO_4$) like $CaWO_4$ and $PbWO_4$ are used as solid state scintillators **[2, 3]**, and there is much interest in the use of scheelite compounds in optoelectronic devices **[4 - 6]**. Moreover, a family of superhard materials has been found in $ABO_4$ compounds with A and B atoms having valence +4 **[7]**.

In the last years there has arisen renewed interest in $ABX_4$ compounds and its evolution under pressure. Many of these compounds crystallize in the scheelite structure (space group: $I4_1/a$, No. 88, Z = 4) or in related structures like zircon (space group: $I4_1/amd$, No. 141, Z = 4), pseudoscheelite (space group: P*nma*, No. 62, Z = 4), wolframite (space group: $P2/c$, No. 13, Z = 2), M-fergusonite (space group: $I2/a$, No. 15, Z = 4), hereafter called fergusonite, and M'-fergusonite (space group: $P2_1/c$, No. 14, Z = 2). In particular, the ambient conditions scheelite structure of $CaWO_4$ and $SrWO_4$ has eight symmetry elements and a body-centered tetragonal primitive cell that includes two formula units, see **Fig. 1(a)**. Each W site is surrounded by four equivalent O sites in tetrahedral symmetry about that site. Each Ca (Sr) cation shares corners with eight adjacent $WO_4$ tetrahedra.

Several experimental and theoretical works have been reported in the last decade on the pressure behavior of scheelite oxides and fluorides **[8 - 33]**. Upon compression most of these compounds undergo structural transitions to monoclinic structures. However, several of these low-symmetry structures are difficult to characterize in high-pressure x-ray diffraction experiments and it has been further suggested that their formation could depend on the stress conditions in the pressure chamber. In particular, a discussion regarding the high-pressure phase of $CaWO_4$ was open in recent years **[8, 9]**.

The occurrence of pressure-driven phase transitions in $CaWO_4$ and $CaMoO_4$ was first reported by Nicol and Durana **[10]**, who postulated that the high-pressure phases had the wolframite structure. Other monoclinic structures that were considered during decades as candidate structures for the $ABO_4$ compounds at high pressure were those of $\alpha$-$MnMoO_4$-type (space group: C2/m, No. 12, Z = 8) **[11]**, $BaWO_4$(II)-type (space group: $P2_1/n$, No. 14, Z = 8) **[12]**, and $HgWO_4$-type (space group: C2/c, No. 15, Z = 4) **[13]**. Errandonea and coworkers **[8]** performed for the first time energy-dispersive x-ray powder diffraction (EDXRD) experiments on $CaWO_4$ up to pressures where the high-pressure phase was observed. They observed the occurrence of the pressure-driven phase transition at 10 GPa. These authors considered the four monoclinic structures previously postulated for the high-pressure phase of $CaWO_4$ to index their EDXRD patterns. Based on the quality of the unit-cell fit, they concluded that the high-pressure phase of $CaWO_4$ was most likely of the wolframite-type **[8]** (see **Fig. 1(b)**). The same was also concluded by Shieh *et al.* from a high-pressure x-ray diffraction study on $CdMoO_4$ **[14]**. However, most recently Grzechnik et al. **[9]** performed high-resolution angle-dispersive x-ray powder diffraction (ADXRD) on $CaWO_4$ and reported the high-pressure structure to be fergusonite-like (see **Fig. 1(c)**). Later measurements on $BaWO_4$ **[13]**, $BaMoO_4$ **[33]**, and $CaMoO_4$ **[29]** also reported a scheelite-to-fergusonite phase transition, but in the case of $SrWO_4$, a recent study combining x-ray diffraction and absorption observed a phase transition at 11.7 GPa and characterized the high-pressure phase as wolframite **[32]**. From the theoretical side, support to the scheelite-to-fergusonite transition with increasing pressure in $ABX_4$ scheelite compounds has been given by the works of Sen *et al.* **[15, 16]**, while support to the scheelite-to-wolframite transition was reported in the work of Li *et al.* **[17]**.

In this work we report new high-pressure ADXRD experiments up to nearly 18

GPa and x-ray absorption near-edge structure (XANES) measurements up to nearly 20 GPa on $CaWO_4$ and $SrWO_4$ along with *ab initio* total energy calculations in both compounds. From our ADXRD data we find that under hydrostatic conditions both compounds undergo a scheelite-to-fergusonite phase transition with increasing pressure, which is supported by the high-pressure XANES measurements and the *ab initio* total energy calculations.

**II. Experimental Details**

$CaWO_4$ and $SrWO_4$ crystals were grown with the Czochralski method starting from raw powders having 5N purity **[4]**. Samples were prepared as finely ground powders from the single crystals of $CaWO_4$ and $SrWO_4$. High-pressure ADXRD measurements were carried out in 450 μm culet Merrill-Basset diamond-anvil cell (DAC) for $CaWO_4$ and in a 400 μm culet Mao-Bell DAC for $SrWO_4$. In the first case, powder samples were loaded together with a ruby chip into a 180 μm diameter hole drilled on a 200 μm thick rhenium (Re) gasket pre-indented to 60 μm. In the second case, the Re gaskets were pre-indented to 40 μm and the diameter of the gasket hole was 100 μm. Silicone oil was used as pressure-transmitting medium in both cases. For XANES measurements under pressure, fine powder samples were loaded together with a ruby chip into a 200 μm diameter hole drilled on a 200 μm thick Inconel gasket pre-indented to 50 μm and inserted between the diamonds of a 400 μm culet membrane-type DAC with silicone oil as pressure-transmitting medium. The pressure was measured by the shift of the R1 photoluminescence line of ruby **[34]**.

ADXRD experiments were performed at the 16-IDB beamline of the HPCAT facility at the Advanced Photon Source (APS) using monochromatic radiation with $\lambda$ = 0.3679 Å (a Si (311) double-crystal monochromator was used). The monochromatic x-ray beam was focused down to 10 x 10 $\mu m^2$ using multilayer bimorph mirrors in a Kickpatrick-Baez configuration **[35]**. Diffraction images were recorded with a Mar345 image plate detector, 230 mm away from the sample, and were integrated and corrected for distortions using the FIT2D software **[36]**. The indexing, structure solution, and refinements were performed using the GSAS **[37]** and the POWDERCELL **[38]** program packages.

XANES experiments were conducted at the ID24 energy dispersive x-ray absorption station of the European Synchrotron Radiation Facility (ESRF) **[39, 40]**. The key component of the dispersive setup is a curved monochromator that selects an energy span around the absorption edge and focuses the beam in the horizontal direction. All the energies contained in the diffracted beam are detected simultaneously by means of a position sensitive detector. In order to establish the energy-pixel correlation, the spectrum of a reference standard is measured and compared with an equivalent spectrum acquired with the classical setup, where the knowledge of the Bragg angle allows for a determination of the energy. A more detailed description of the principles of energy-dispersive x-ray-absorption data collection is given in Ref. [**41]**.

All XANES experiments were performed at the W $L_3$-edge (10.207 keV). At ID24, the combination of a profiled curved Si (111) monochromator **[42]** and a vertically focusing mirror defined a focus spot of approximately 30 x 20 $\mu m^2$. The membrane DAC was situated at the focus position. The incident and transmitted beams were alternatively measured. In our experiments, the incident intensity was measured outside the pressure chamber. An essential experimental aspect of x-ray absorption spectroscopy (XAS) experiments in a DAC is the presence of diffraction peaks originating from diffraction from the diamond single crystals. The pressure cell is oriented with respect to the polychromatic x-ray beam so as to remove these glitches from the widest spectral range around the x-ray-absorption edge. This operation takes advantage of the real time visualization of the XAS spectra, characteristic of the energy-dispersive setup. The presence of harmonics was avoided thanks to the grazing incidence mirrors situated between the undulator source and the monochromator. The reference standard for the energy calibration was metallic W.

## III. Overview of the calculations

The structural stability of the phases of $CaWO_4$ and $SrWO_4$ was further investigated theoretically by means of total energy calculations performed within the framework of the density functional theory (DFT) with the Vienna *Ab Initio* Simulation Package (VASP) **[43]**. A review of DFT-based total energy-methods as applied to the theoretical study of phase stability can be found in Ref. **[44]**. The exchange and correlation energy was evaluated within the generalized gradient approximation (GGA) **[45]**. We used ultrasoft Vanderbilt-type pseudopotentials **[46]** and basis sets including plane waves up to a kinetic-energy cutoff of 850 eV for $CaWO_4$ and 495 eV for $SrWO_4$. The tetrahedron method combined with Blöchl corrections was used for the Brillouin-zone integrations. The total energies were converged to below 1 meV per formula unit. The structural relaxation of the phases at each volume was conducted through the calculation of the forces on the atoms and the components of the stress tensor.

## IV. Results and discussion

### A. ADXRD measurements at high pressures

### A. 1. Low-pressure phase

**Fig. 2** shows our ADXRD data for $CaWO_4$ and $SrWO_4$ at several selected pressures up to 18 GPa. The evolution with pressure of the volume, lattice parameters, and axial ratios is plotted in **Figs. 3 and 4** where we also compare our results with previously reported data for $CaWO_4$ **[8, 9, 19, 47, 48]** and $SrWO_4$ **[47, 49]** (in this case only for ambient pressure).

The pressure-volume (P-V) curves shown in **Fig. 3(a)** were analyzed in the standard way using a Birch-Murnaghan equation of states (EOS) **[50]**,

$$P = \frac{3}{2} B_0 (x^{7/3} - x^{5/3})[1 + \frac{3}{4}(B_0^{'} - 4)(x^{2/3} - 1)], \qquad (1)$$

with $x = V_0 / V$, where the parameters $V_0$, $B_0$, and $B_0$' are the zero-pressure volume, bulk modulus, and pressure derivative of the bulk modulus, respectively. For scheelite $CaWO_4$ we find $V_0$ = 312(1) Å$^3$, $B_0$ = 74(7) GPa, and $B_0'$ = 5.6(9) [$V_0$ = 347.4(9) Å$^3$, $B_0$ = 63(7) GPa, and $B_0'$ = 5.2(9) for scheelite $SrWO_4$]. These parameters are in good agreement with previous reported results **[9, 19]** and indicate that $SrWO_4$ is more compressible than $CaWO_4$, which is a direct consequence of the different compressibility of the *c*-axis in the two compounds, see below. It is worth to mention that the evolution of the volume of $CaWO_4$ with pressure reported in **Ref. [8],** and plotted as solid squares in **Fig. 3(a)** for the sake of comparison, underestimates the decrease of the volume above 7 GPa. This result gives support to the idea that a non-hydrostatic pressure environment may affect the structural pressure behavior of scheelite tungstates, as we will comment later on.

**Fig. 3(b)** shows that the compressibility of the *c*-axis of the scheelite structure is larger for $SrWO_4$ than for $CaWO_4$, while the *a*-axis compresses in the same way in the two compounds (see **Fig. 3(c)**). The larger compressibility of the *c*-axis in $SrWO_4$ compared to that of $CaWO_4$ can be related to the difference in size of the $Ca^{+2}$ and $Sr^{+2}$ cations, which implies a larger charge density in the Ca environment with respect to that around Sr, as we will discuss later. The larger compressibility along the *c*-axis as compared to that along the *a*-axis is evident in **Fig. 4**.

We have also investigated the evolution of cation-anion distances in both compounds. According to the single-crystal high-pressure investigation carried out by Hazen *et al.* **[19]** up to 4.1 GPa, the relative positions of the atoms in the $CaWO_4$ unit cell do not vary under pressure within the experimental error. In our experiment we have determined the internal parameters at the lowest pressure by means of a Rietveld refinement and then maintained them constant at higher pressures (see Table I). **Fig. 5** shows the evolution of the atomic distances between nearest neighbors with increasing pressure. The interatomic distances in $CaWO_4$ evolve in a similar way as previously reported **[19, 27],** but the present results systematically differ by less than ~2% from those reported in Ref. **[27]**. This difference was observed before by Hazen **[19]** between experiments performed inside and outside a DAC and can be attributed to the limited access to the reciprocal space of the used DAC **[19]** and to the presence of impurities in the studied samples **[51]**. The good agreement between our results and previous ambient pressure results **[19, 52]** suggests that the pressure evolution of the interatomic distances reported here is more reliable than previous published data. The decrease of Ca-O and Sr-O distances can be compared with the rigidity of the W-O bond distance in both compounds. In **Fig. 5** it can be seen that there are two Ca-O and Sr-O distances, the largest distances being more compressible than the shorter ones.

Our results support the description of $AWO_4$ tungstate scheelites in terms of hard anion-like $WO_4$ tetrahedra surrounded by charge compensating cations. When pressure is applied the $WO_4$ units remain essentially undistorted and the reduction of the unit-cell size is mainly associated to the compression of the A cation polyhedral environment **[19]**. Along the *a*-axis the $WO_4$ units are directly aligned, whereas along the *c*-axis there is an A cation between two $WO_4$ tetrahedra. Therefore, the different arrangement of hard $WO_4$ tetrahedra along the *c*- and *a*-axis accounts for the different compressibility of the two cell axes. The different pressure behavior of the two A-O distances (**Fig. 5**) is associated to the different compressibility of the cell parameters. Effectively, the longest A-O distance has the largest projection along the *c*-axis.  It is important to point out that the asymmetric behavior of *c*- and *a*-axis is also revealed in their different thermal expansion **[53]**, as well as in the evolution of the *c/a* ratio along a cationic A series **[47]**.

**A.2. High-pressure phases**

The ADXRD spectra of $CaWO_4$ exhibit a change around 11.3 GPa, while in $SrWO_4$ the change occurs near 10.1 GPa (see **Fig. 2**). These changes are completely reversible upon pressure release. Below those pressures the observed diffraction peaks shift smoothly with compression and all the reflections observed in the diffraction patterns can be indexed within the scheelite structure whereas above those pressures some of the diffraction peaks split and additional diffraction peaks emerge. In particular, the appearance of a new peak around $2\theta \approx 3.8^o$ (depicted by an arrow in **Fig. 2**) is clearly distinguishable**.** The observed splitting of peaks and the appearance of new reflections suggests the occurrence of a second-order phase transition. The measured ADXRD patterns of the high-pressure phase can be indexed on the basis of the fergusonite structure but not on the basis of the wolframite structure, confirming Grzechnik's results for $CaWO_4$ **[9]**. The new Bragg peaks observed at $2\theta \approx 3.8^o$ in the high-pressure phase of both compounds correspond to the (020) reflection of the fergusonite structure of $CaWO_4$ and $SrWO_4$. Two further facts support the assignment of the fergusonite structure to the high-pressure phase of both compounds and rule out the wolframite structure: The first one is that two of the stronger Bragg peaks of the wolframite structure, viz. the (011) and (110) expected at $2\theta \approx 5.7^o$, are absent in the measured diffraction patterns. The second one is that the (100) reflection of the wolframite structure is not present at $2\theta \approx 4.15^o$.

**Fig. 2** also shows the Rietveld refinements to the experimental spectra of $CaWO_4$ at 11.3 GPa and of $SrWO_4$ at 10.1 GPa obtained assuming the fergusonite structure. In order to perform the Rietveld refinement the starting Ca (Sr), W, and O positions were taken from Ref. **[9]**. For both tungstates, we obtained good agreement with the experimental diffraction patterns. The residuals are: $R_{WP}$ = 1.75%, $R_P$ = 1.1%, and $R(F^2)$ = 1.5% for $CaWO_4$ (197 reflections) and $R_{WP}$ = 2.07%, $R_P$ = 1.4%, and $R(F^2)$ = 1.9% for $SrWO_4$ (324 reflections). Similar refinement quality was obtained for scheelite $CaWO_4$ at 1.4 GPa and scheelite $SrWO_4$ at 0.2 GPa. Table I summarizes the lattice parameters and atomic positions of $CaWO_4$ at 1.4 and 11.3 GPa, and of $SrWO_4$ at 0.2 and 10.1 GPa. Our structural parameters for fergusonite $CaWO_4$ agree with those reported by Grzechnik *et al.* **[9]**.

It is worthwhile to discuss here the differences between the present and Grzechnik´s results **[9]** with previous structural studies on $CaWO_4$ and $SrWO_4$. As we mentioned above, in a previous EDXRD study Errandonea *et al.* **[8]** characterized the high-pressure phase of $CaWO_4$ as wolframite-type. This conclusion was a result of a LeBail analysis **[54]** considering four candidate structures, among which the fergusonite structure was not included. The exclusion of this structure was not accidental but a consequence of the fact that the (020) Bragg peak and other characteristic reflections of the fergusonite structure were not present in the EDXRD patterns of the high-pressure phase reported in Ref. **[8]**. Furthermore, in these patterns there are also two reflections around 23 keV which were assigned to the (011) and (110) Bragg peaks of the wolframite structure and which cannot be indexed with the fergusonite structure – that is, the experimental situation was quite different from what we observe in the present experiments. We think that in the previous EDXRD experiments the presence of large non-hydrostatic stresses inside the DAC **[8]** may have favored a transition to the wolframite structure instead of the fergusonite structure. In Grzechnik's study, both helium and a 4:1 methanol-ethanol mixture were used as pressure-transmitting medium **[9]**. In the present study silicone oil was used as pressure transmitting medium. In contrast, in Ref. **[8]** no pressure-transmitting medium was used. Using a non-hydrostatic pressure medium as NaCl, Nicol and Durana assigned the wolframite structure to the high-pressure phase of $CaWO_4$ **[10].** The bulk modulus of $CaWO_4$ is three times larger than that of NaCl and therefore the absence of a pressure-transmitting medium could create highly non-hydrostatic conditions at the onset of the transition **[55]**. It is well known that phase transitions can be greatly affected by non-hydrostatic conditions **[55]** and therefore the fact that the less hydrostatic media was used in Ref. **[8]** could then have affected the characterization of the high-pressure phase of $CaWO_4$. The observation of a scheelite-to-wolframite transition in $CdMoO_4$ in experiments performed by Shieh *et al.* **[14]** using $CdMoO_4$ without pressure-transmitting medium, as well as the differences between the compressibility observed for the scheelite phase in these experiments and the one observed when a 4:1 methanol-ethanol mixture was used as pressure-transmitting medium **[19]**, give additional support to this hypothesis. Regarding $SrWO_4$, Kuzmin *et al.* **[32]** concluded recently from their x-ray diffraction and absorption measurements that the high-pressure phase of this compound is of the wolframite-type. There are two principal facts that may explain the differences between the results reported in Ref. **[32]** and the present results. The first one is that the lower quality of the EDXRD patterns reported in Ref. **[32]** in comparison with the ADXRD patterns reported here. The x-ray patterns reported in Ref. **[32]** do not allow the authors to perform a structural refinement and the only they can conclude is that there is a phase transition at 11.7 GPa, a pressure that is in fairly good agreement with our own results. The second one is that the Extended X-ray Absorption Fine Structure (EXAFS) measurements reported in Ref. **[32]** show that the local structure around the W atoms is compatible with an octahedral coordination at 30 GPa. However, from the EXAFS analysis alone, it is not possible to identify the structure of the high-pressure phase.

Then, the possible existence of a post-fergusonite phase with the tungsten atoms in an octahedral coordination will resolve apparent controversies between our results and those reported by Kuzmin *et al.* **[32]**. Another fact to be taken into consideration is the possible metastability of two different monoclinic structures, an scenario that is supported by the polytypism observed in other tungstates (e.g. $PbWO_4$) even at ambient conditions **[56]**.

**Figs. 3(b) and 3(c)** shows the lattice parameters of the fergusonite phases of $CaWO_4$ and $SrWO_4$ as a function of pressure up to ~18 GPa. Above 15 GPa the quality of the ADXRD patterns deteriorated, but it was still possible to obtain the lattice parameters at different pressures using the LeBail extraction technique **[54]**. The degradation of the x-ray diffraction patterns was observed previously in $CaWO_4$ **[9]** and in similar compounds **[13, 57]**, and is independent of the pressure-transmitting medium employed in the experiments. This observation may be related to precursor effects either of a martensitic transition **[58]** or of the amorphization observed in alkaline-earth tungstates **[8]** and other scheelite-structured compounds **[31]** at higher pressures. The $\beta$ angle was found to increase slightly from 90.09° at 11.3 GPa to 93° at 18.3 GPa in $CaWO_4$ and from 90.35° at 10.1 GPa to 92° at 17.5 GPa in $SrWO_4$. The difference between the *b/a* and *b/c* axial ratios of the fergusonite phases of $CaWO_4$ and $SrWO_4$ also increases upon compression, see **Fig. 4**. These two facts imply an increase of the monoclinic distortion with pressure. A volume discontinuity is not apparent at the transition pressure, consistent with a second-order phase transition. The Birch-Murnaghan fit to both the scheelite and the fergusonite pressure-volume data gives EOS parameters ($V_0$, $B_0$, and $B_0'$) that differ by less than one standard deviation from those obtained for the scheelite data only. Hence, the EOS reported above can be assumed as a valid EOS for $CaWO_4$ and $SrWO_4$ up to 18 GPa, as illustrated in **Fig. 3(a)**. A Birch-Murnaghan fit to only the high-pressure fergusonite data gives slightly larger values for $B_0$ and $B_0'$ [e.g. for $CaWO_4$ we obtained: $V_0$ = 312(2) Å$^3$, $B_0$ = 78(9) GPa, and $B_0'$ = 5.7(12) and for $SrWO_4$: $V_0$ = 347(2) Å$^3$, $B_0$ = 64(8) GPa, and $B_0'$ = 5.4(11)]. A similar conclusion can be drawn from our *ab initio* calculations, see Sec. IV.C.

In order to close the discussion on the ADXRD results we would like to comment that in both compounds the phase transition implies a distortion of the $WO_4$ tetrahedra accompanied by a small shear distortion of alternate (100) cation planes in the [001] direction. The scheelite-to-fergusonite transition occurs together with a slight decrease of two W-O bonds and the increase of the other two W-O bonds inside the $WO_4$ tetrahedra, however, as a consequence of this deformation, the volume of the $WO_4$ tetrahedra is enlarged less than 10%. On the other hand, at the transition six of the A-O bonds in the $AO_8$ polyhedra are compressed and the remaining two are enormously expanded, see **Fig. 5**. The consequence of these changes is a decrease of the volume of the $AO_8$ polyhedra. In this way, as a result of the phase transition the $WO_4$ tetrahedra in the fergusonite phase are only slightly distorted, while the $AO_8$ polyhedra are quite distorted (see **Fig. 1**).

### B. XANES measurements at high pressures

#### B.1. Low-pressure phase

The XANES part of the absorption spectrum is very sensitive to modifications in the neighborhood of the absorbing atom and thus it can be used as a tool to detect structural changes. We have performed XANES experiments on $CaWO_4$ and $SrWO_4$ under compression with the aim of investigating changes in W coordination after the phase transition. In the scheelite structure the W environment is formed by four O atoms in tetrahedral configuration. If the high-pressure phase were fergusonite, the tetrahedron would become distorted which results in two slightly different near-neighbor distances but the main characteristics of the W environment would be maintained. In this situation we would expect small changes in the XANES spectra. If however the high-pressure phase would be of the wolframite-type the W coordination would change to six (2+4) and one would expect significant changes in the XANES spectra.

In order to confirm these ideas and as a guide to interpret changes in the experimental spectra, we have performed XANES simulations of the scheelite, fergusonite and wolframite phases. The XANES simulations were carried out using the real-space multiple-scattering code implemented in the FEFF8 package **[59]**. We employed a self-consistent potential calculated using 120 atoms clusters (6.9 Å or 14 shells) and the Hedin-Lundqvist energy-dependent self-energy. Full multiple-scattering XANES calculations were performed using 87-atom clusters (6.5 Å or 11 shells). No pseudo Debye-Waller factor has been considered in our simulations. The structural data used are given in Table I for the scheelite and fergusonite structures and in Table II for the wolframite structure. The description of wolframite is based on that of $CdWO_4$ **[60]**. For this structure, the lattice parameters have been scaled to give the same volume per formula unit as in the fergusonite structure. In **Fig. 6** we present the results for the XANES spectra simulated in the three structures for $CaWO_4$ and $SrWO_4$. The spectra corresponding to both compounds are similar, with five resonances. The most dramatic change observed when passing from fourfold coordination to sixfold coordination affects the resonance named B in **Fig. 6**. In the scheelite and fergusonite structures the B resonance is clearly observable, but it disappears in the wolframite simulation. Other noticeable changes concern the intensity and width of the white line (A resonance).

**Fig. 7** shows the experimental XANES spectra at different pressures for $CaWO_4$ and $SrWO_4$. The spectra of both compounds at atmospheric pressure show the five resonances predicted by our simulations for the scheelite structure. The position and intensity of each feature agree qualitatively with those of the simulation, except for the resonances D and E in $CaWO_4$ whose relative intensities are inverted. In the theoretical spectra the resonances are more pronounced as a consequence of not considering the pseudo Debye-Waller factor.

**B.2. High-pressure phases**

The high pressure XANES spectra of $CaWO_4$ show no significant changes up to 11.3 GPa, see **Fig. 7(a)**. At this pressure the B resonance looses intensity and the ratio of intensities between the D and E resonances also decreases. Meanwhile the intensity and width of the white line remain unaffected. The changes described indicate a transition to the fergusonite phase at 11.3(10) GPa, in agreement with ADXRD results. It is interesting to note that XANES spectra continue to evolve up to the maximum pressure attained of 20.2 GPa suggesting, as we observed in our ADXRD measurements, that the structural distortions leading to the fergusonite structure become more pronounced when applying pressure. The phase transition is reversible, as the spectrum of the recovered phase is identical to the initial one except for a diminution in the white line intensity which we interpret as due to a decrease in sample thickness.

As regards to $SrWO_4$ the XANES spectra up to 12.4 GPa show only a small reduction of the intensity of the B resonance, see **Fig. 7(b)**. At 15.0 GPa an acceleration in the decrease of the B resonance is accompanied by the progressive disappearance of the C resonance and an increase of the D resonance, while the white line remains unchanged. These changes continue up to the maximum pressure attained of 22.2 GPa. At this pressure the B resonance is still visible in the spectrum. Once again, the evolution of the spectra is reversible and suggests a transition towards the fergusonite phase. However, the onset of the phase transition is not as clear as in $CaWO_4$ and the distortion of the W tetrahedral environment is not evident up to 13.7(17) GPa.

### C. Ab initio calculations

We compare now the experimental body of data presented in the previous sections with the results from our total-energy theoretical study of several structural phases of $CaWO_4$ and $SrWO_4$. Along with the observed scheelite and fergusonite phases we have also considered the wolframite structure previously proposed for the high-pressure phase of $CaWO_4$ **[8]** as well as other candidate structures on account of their observation or postulation in previous high-pressure work for related compounds: M'-fergusonite **[16]**, $LaTaO_4$ **[61]**, $BaWO_4$-II **[12]**, and $YLiF_4$-Sen (as we call the very-high-pressure structure found in the molecular dynamics study reported by Sen *et al.* **[16]**). Several of these phases are structurally related and can be represented within the monoclinic space group $P2_1/c$ (No. 14), which has thus also received our special attention.

**Fig. 8** shows the energy-volume curves for the different structures of $CaWO_4$ and $SrWO_4$, from which the relative stability and coexistence pressures of the phases can be extracted by the common-tangent construction **[44]**. At all the pressures investigated and for both compounds the M'-fergusonite structure reduced upon full relaxation to fergusonite – it is thus not shown in **Fig. 8**. This figure shows the scheelite phases as being stable at zero and low pressure, with $V_0$ = 318.3 Å$^3$, $B_0$ = 72 GPa, and $B_0'$ = 4.3 for $CaWO_4$ and $V_0$ = 362.2 Å$^3$, $B_0$ = 62 GPa, and $B_0'$ = 4.9 for $SrWO_4$. These values compare well with the experimental results, with differences within the typical reported systematic errors in DFT-GGA calculations. A similar degree of agreement exists for the calculated values of the internal parameters of the scheelite phases [O(16f) at (0.244, 0.097, 0. 039) and *c/a* = 2.16 for $CaWO_4$; O(16f) at (0.237, 0.111, 0.042) and *c/a* = 2.20 for $SrWO_4$, cf. Table I].

As pressure increases, the scheelite structure becomes unstable against fergusonite. The fergusonite structure, a distortion of scheelite, only emerges as a structurally different and thermodynamically stable phase above a compression threshold of about 10 - 11 GPa in both compounds; at the lower pressures investigated the relaxation of the fergusonite structure resulted in the scheelite structure. This is consistent with a continuous or quasi-continuous scheelite-to-fergusonite transition with none or very little volume collapse. The calculated structural parameters of the fergusonite phases are also in good agreement with the experimental results [*y*(Ca)= 0.624, *y*(W) = 0.132, $O_1$(8f) at (0.912, 0.963, 0.242), $O_2$(8f) at (0.492, 0.217, 0.822), *b/a* = 2.104, *c/a*= 0.977, and β= 91.6 for $CaWO_4$ at 11 GPa; *y*(Sr)= 0.624, *y*(W) = 0.128, $O_1$(8f) at (0.905, 0.961, 0.235), $O_2$(8f) at (0.485, 0.213, 0.840), *b/a* = 2.145, *c/a*= 0.990, and β= 90.3 for $SrWO_4$ at 11 GPa, cf. Table I].

The $BaWO_4$-II and $YLiF_4$-Sen structures are very high in enthalpy and nowhere close to stability in either compound. The $LaTaO_4$-type structure is similarly high in enthalpy in $CaWO_4$ though in $SrWO_4$ it is placed considerably lower and is in fact a competitive candidate for stability in a post-fergusonite regime around 20 GPa. The wolframite structure is not thermodynamically stable in any interval of pressures though it is close in energy (20 - 40 meV) to fergusonite in $CaWO_4$ in the relevant range around 10 - 20 GPa which might have a bearing on its observation in previous experimental work in which non-hydrostatic conditions were used **[8]**.

A difficulty found in the relaxation of the monoclinic phases belonging to space group $P2_1/c$ is the existence of a number of local minima. For a significant interval of medium and high pressures these structurally different minima are located very close in energy, sometimes separated by shallow barriers, which make the precise determination of the absolute minimum within this set of low-symmetry crystal structures a rather tedious and difficult task. Nevertheless we have carried out such minimization ensuring great care in the relaxation procedure, which requires in particular repeating relaxation starting from different initial conditions and checking for local stability. In the course of this minimum-trapping quest we have arrived at a well defined minimum in the compressed region for a structure which after refinement and further analysis turned out to have *increased* orthorhombic symmetry, with space group *Cmca* (No. 64). This totally unexpected structural phase **[62]** has lower enthalpy than any other of the phases considered above ~29 GPa in $CaWO_4$ and ~21 GPa in $SrWO_4$ (in this case in close competition with the $LaTaO_4$-type structure –see **Fig. 8(b)**). It has Z = 8 with Ca atoms in 8e positions at (0.25, 0.164, 0.25), W in 8f (0, 0.409, 0.226) and O at 8d (0.157, 0.5, 0), 8e (0.25, 0.348, 0.25), 8f (0, 0.288, 0.005), and 8f (0, 0.084, 0.094) for $CaWO_4$ at 30 GPa [for $SrWO_4$ at 23 GPa: Sr(8e) (0.25, 0.167, 0.25), W(8f) (0, 0.413, 0.223), $O_1$(8d) (0.149, 0.5, 0), $O_2$(8e) (0.25, 0.359, 0.25), $O_3$(8f) (0, 0.292, 0.034), and $O_4$(8f) (0, 0.084, 0.077)]. In both materials *b/a*~1.65 - 1.68 and *c/a*~0.68. In this structure the Ca (Sr) and W cations are surrounded by 10 and 6 O atoms, respectively. It is worth noting that this new structure is *strongly* energetically favored over fergusonite in the high-pressure regime and thus the figures of ~29 GPa in $CaWO_4$ and ~21 GPa in $SrWO_4$ constitute neat upper bounds for the thermodynamical stability of the respective fergusonite phases. Such high pressures are just above those reached in x-ray diffraction experiments.

### D. Bulk modulus in scheelite $ABO_4$ compounds.

Hazen *et al.* found that the bulk modulus of certain binary oxides and silicates can be directly correlated to the compressibility of the A cation coordination polyhedra **[63]**. In particular, they proposed that the bulk compressibility in these compounds is proportional to the average volume of the cation polyhedron divided by the cation formal charge; i.e., $B_0$ is proportional to the cation charge density per unit volume inside the cation polyhedron. They also found that $A^{2+}B^{6+}O_4$ scheelite tungstates and molybdates under pressure compressed in an anisotropic way with the $WO_4$ and $MoO_4$ tetrahedra behaving as rigid units **[19]**. Furthermore, they ordered the compressibility of scheelite compounds according to the A cation formal charge and, on this basis, suggested that the compressibility of $ABO_4$ scheelites could be given by the compressibility of the softer $AO_8$ polyhedron and that the $A^{4+}B^{4+}O_4$ scheelites could be a family of ultrahard materials.

These last conclusions have been confirmed in two recent works, where the bulk moduli of scheelites have been plotted as a function of the bulk volume **[7, 64]**. A further insight can be obtained with the present data by plotting the bulk modulus of scheelite and scheelite-related compounds as a function of the A cation charge density per unit volume in the $AO_8$ polyhedra, given by the A cation formal charge divided by the cubic average A-O distance (see **Fig. 9**). All data plotted in **Fig. 9,** summarized in Table III, correspond to approximately 25% of the $ABO_4$ compounds with the scheelite and scheelite-related structures that can be found in the Inorganic Crystal Structure Database. The bulk modulus of all the plotted compounds obeys a linear relationship according to the equation:

$$B_0 = 610(110)\frac{Z_A}{d_{A-O}^3} \quad , \qquad (2)$$

where $B_0$ is the bulk modulus (in GPa), $Z_A$ is the A cation formal charge (being $4 \geq Z_A \geq 1$), and $d_{A\text{-}O}$ is the average A-O distance (in Å) inside the $AO_8$ polyhedron. This simple rule serves as an effective and simple empirical criterion for predicting the bulk modulus of any scheelite or scheelite-related $ABO_4$ compound. The linear relationship between $B_0$ and the A cation charge density of the $AO_8$ polyhedra is consistent with the fact that $AO_8$ polyhedra exhibiting a large A cation charge density result in a larger electronic cloud inside the polyhedra than those $AO_8$ polyhedra with a low A cation charge density. In the $AO_8$ polyhedra with a high $Z_A$ the electrons around the cation are highly localized and the bond distances cannot be highly deformed under pressure. On the contrary, in $AO_8$ polyhedra with a low $Z_A$ the density electrons around the cation are highly delocalized and the bond distances can be considerably deformed under pressure. Then, since the compressibility of $ABO_4$ compounds is mainly given by the compression of the $AO_8$ polyhedra, the above described facts explains why $B_0$ is proportional to $Z_A$. In addition to that, they also explains why $AO_8$ polyhedra with A valence (+1, +2, and +3) are highly deformed as compared to $BO_4$ polyhedra with B valence (+7, +6, and +5) in $ABO_4$ scheelites and scheelite-related structures, being the compounds with A and B cation valence equal to +4 the hardest $ABO_4$ materials. In fact, the linear relationship stated above should not be applicable to $A^{4+}B^{4+}O_4$ scheelites if $AO_8$ and $BO_4$ tetrahedra have similar compressibilities. However, despite both A and B cations having equal valence, B-O bonds in tetrahedral configuration are shorter and stronger than A-O bonds and the bulk modulus is again dictated by the $AO_8$ polyhedra. Therefore, Eq. (2) can also be effectively applied to $A^{4+}B^{4+}O_4$ scheelites as clearly shown.

It has been recently reported that both the scheelite and the zircon structure of $YVO_4$ have a quite similar bulk modulus **[64]**. This result is in agreement with our expectations since in both structures the Y-O bond distances differ by less than 2%. A similar behavior should be expected also for $ZrSiO_4$, with similar Zr-O bond distances in the scheelite and zircon structures (see Table III). However, a bulk modulus of 300 GPa has been recently reported for the scheelite phase of $ZrSiO_4$ **[7]**. This bulk modulus exceeds by more than 30% the bulk moduli of the zircon structure of $ZrSiO_4$. Therefore, according to the systematic here reported, a bulk modulus of 300 GPa for the scheelite phase of $ZrSiO_4$ is unrealistic and we think that the extremely low compressibility recently reported for this material could be mistaken. Following Eq. (2), we can predict for the scheelite phase of $ZrSiO_4$ a bulk modulus of 220(40) GPa, which is one of the largest bulk modulus of $ABO_4$ compounds. Theoretical calculations using either the local-density approximation (LDA) or the generalized-gradient approximation (GGA) gave a bulk modulus of 230(25) GPa **[81]**, value that agrees well with our estimation. A bulk modulus of 300 GPa can be only expected for a compound with octahedral coordinated silicon atoms, like $\gamma$-$Si_3N_4$, but not for compounds with tetrahedral coordinated Si atoms **[82]**, like scheelite $ZrSiO_4$. We attributed the overestimation of $B_0$ to: i) the non-hydrostatic conditions of the experiments performed by Scott *et al.* **[7]** who used a 16:3:1 methanol-ethanol-water mixture as pressure-transmitting medium up to 52.5 GPa, and ii) to the large presence of impurities in the natural zircon samples used by Scott *et al.*, as suggested by Van Westrenen *et al.* **[51]**. The first argument leads to large pressure gradients and inaccurate estimation of the pressure inside the DAC when a 60 μm x-ray beam is used because the pressure transmitting medium used is not hydrostatic above 15 GPa. In fact, the Pt diffraction peaks used for determining the pressure in Ref. **[7]** are quite broad. These facts may easily cause an overestimation of the bulk modulus of the scheelite phase of $ZrSiO_4$. The second argument has proved to lead to different transition pressures and different pressure coefficients. New experiments using a micro-focus x-ray beam and better hydrostatic conditions are needed to check the pressure behavior of the scheelite phase of $ZrSiO_4$.

To conclude, we would like to mention that attempting to predict the pressure behavior of other scheelite-structures and zircon-structured $ABO_4$ materials we used Eq. (2) to make a back-of-the-envelope estimation of the bulk modulus of several compounds, which have been selected by considering their actual technological interest.

Our predictions are summarized in Table IV. In the case of $BaMoO_4$, our estimation of $B_0$ is in quite good agreement with the recent experimental results of Panchal *et al.* **[33]**. On top of that, according to our estimations, hafnon ($HfSiO_4$) is expected to be one of the least compressible $ABO_4$ compounds, being therefore a material of interest for potential applications as an interphase component in toughened oxide ceramic composites **[83]**. Our predictions for $NaReO_4$ can be compared with the bulk modulus obtained from DFT calculations by Spitaler *et al.* **[84]**. These authors reported $B_0$ = 18.3 GPa. This value is approximately half of the value estimated by us. However, a Birch-Murnaghan fit to the results reported by Spitaler *et al.* gives a negative value for the pressure derivative of $B_0$, something unexpected for a scheelite $ABO_4$ compound, which suggests the EOS of $NaReO_4$ may be miscalculated in Ref. **[84]**. This conclusion is also supported by the fact that the value predicted by us for $B_0$ is very similar to that experimentally observed in other perrhenates (see Table III), as expected.

## V. Conclusions

We have measured ADXRD and XANES spectra in $CaWO_4$ and $SrWO_4$ under pressure up to ~20 GPa. In both cases our results support the existence of a reversible scheelite-to-fergusonite structural transition under hydrostatic conditions. From our ADXRD data we locate the onset of the transition at 10.8(5) GPa in $CaWO_4$ and at 9.9(2) GPa in $SrWO_4$. The monoclinic distortion triggered at the phase transition continue up to the maximum pressures attained in our experiment, with no evidence of any further structural transformation. The small changes of the local environment around the absorbing atom make XANES sensitive to the phase transition at slightly higher pressures, around 11.3(10) GPa in $CaWO_4$ and 13.7(17) GPa in $SrWO_4$. In the case of $SrWO_4$ precursor effects of the transition appear at 10 GPa but the transition is not completed up to 15 GPa. The sluggish character of the transition is confirmed not only by the present ADXRD and XANES experiments, but also by the Raman investigation carried out in Ref. **[30]**, where the pressure dependence of some modes related to the internal movement in the $WO_4$ tetrahedra are found to be strongly nonlinear up to 3 - 4 GPa above the transition pressure. Our *ab initio* theoretical study of the energetic of the phases support the scheelite-to-fergusonite transition and yield structural characteristics for the scheelite and fergusonite phases in very good agreement with the experimental results. In addition, from our *ab initio* study we can place an upper bound (not reached experimentally) to the stability of the fergusonite high-pressure phases, at ~29 GPa in $CaWO_4$ and ~21 GPa in $SrWO_4$, which calls for experimental structural studies in this higher pressure region. Finally, we have showed that the ambient-pressure bulk modulus of $ABO_4$ scheelite and scheelite-related compounds can be easily estimated if the average A-O distance is known.

**Acknowledgments**

The authors thank P. Bohacek (Institute of Physics, Prague) for providing the $CaWO_4$ and $SrWO_4$ crystals used in the experiments. This study was made possible through financial support from the Spanish government MCYT under grants MAT2002-04539-CO2-01 and -02, and MAT2004-05867-C03-03 and-01. The U.S. Department of Energy, Office of Science, Office of Basic Energy Sciences supported the use of the Advanced Photon Source (APS) under contract No. W-31-109-Eng-38. DOE-BES, DOE-NNSA, NSF, DOD-TACOM, and the W.M. Keck Foundation supported the use of the HPCAT facility. We would like to thank D. Häusermann and the rest of the staff at the HPCAT of the APS for their contribution to the success of the ADXRD experiments. The XANES experiments were done under proposal number HS-2412 at the ESRF. D. Errandonea acknowledges the financial support from the MCYT of Spain and the Universitat of València through the “Ramón y Cajal” program of grants. J.

López-Solano and A. Muñoz acknowledge the financial support from the Gobierno Autónomo de Canarias (PI2003/174).

**Table I:** Structural parameters of the scheelite and fergusonite structure of $CaWO_4$ and $SrWO_4$. These parameters were obtained from the present Rietveld refinements (see text).

**a)** Structural parameters of scheelite $CaWO_4$ at 1.4 GPa:
$I4_1/a$, Z = 4, $a$ = 5.205(5) Å, $c$ = 11.275(7) Å

| | Site | *x* | *y* | *z* |
|---|---|---|---|---|
| Ca | 4b | 0 | 0.25 | 0.625 |
| W | 4a | 0 | 0.25 | 0.125 |
| O | 16f | 0.2289(3) | 0.0910(4) | 0.0421(5) |

**b)** Structural parameters of fergusonite $CaWO_4$ at 11.3 GPa:
$I2/a$, Z = 4, $a$ = 5.069(2) Å, $b$ = 10.851(5) Å, $c$ = 5.081(7) Å, β = 90.091(9)

| | Site | *x* | *y* | *z* |
|---|---|---|---|---|
| Ca | 4e | 0.25 | 0.6100(8) | 0 |
| W | 4e | 0.25 | 0.1325(3) | 0 |
| $O_1$ | 8f | 0.9309(39) | 0.9684(23) | 0.2421(24) |
| $O_2$ | 8f | 0.4850(35) | 0.2193(31) | 0.8637(37) |

**c)** Structural parameters of scheelite $SrWO_4$ at 0.2 GPa:
$I4_1/a$, Z = 4, $a$ = 5.391(8) Å, $c$ = 11.893(7) Å

| | Site | *x* | *y* | *z* |
|---|---|---|---|---|
| Sr | 4b | 0 | 0.25 | 0.625 |
| W | 4a | 0 | 0.25 | 0.125 |
| O | 16f | 0.2497(9) | 0.0925(9) | 0.0421(6) |

Structural parameters of fergusonite $SrWO_4$ at 10.1 GPa:
$I2/a$, Z = 4, $a$ = 5.263(9) Å, $b$ = 11.182(6) Å, $c$ = 5.231(6) Å, β = 90.35(1)

| | Site | *x* | *y* | *z* |
|---|---|---|---|---|
| Sr | 4e | 0.25 | 0.6027(9) | 0 |
| W | 4e | 0.25 | 0.1243(8) | 0 |
| $O_1$ | 8f | 0.9309(49) | 0.9598(53) | 0.2619(42) |
| $O_2$ | 8f | 0.4903(39) | 0.2278(35) | 0.8779(32) |

**Table II:** Atomic positions used to perform the XANES simulations for the wolframite structure ($P2/c$, Z = 2) **[60]**.

| | Site | *x* | *y* | *z* |
|---|---|---|---|---|
| A | 2f | 0.5 | 0.3027 | 0.75 |
| W | 2e | 0 | 0.1785 | 0. 25 |
| $O_1$ | 4g | 0.242 | 0.372 | 0.384 |
| $O_2$ | 4g | 0.202 | 0.096 | 0.951 |

**Table III:** Summary of the data plotted in **Fig. 9**. The structure, A-O bond distance, cation formal charge, and bulk modulus are given.

| $ABO_4$ compound | Space Group | mean A-O bond distance [Å] | cation formal charge | $B_0$ [GPa] | Reference |
|---|---|---|---|---|---|
| $ZrSiO_4$ | I4$_1$/*a* | 2.243 | 4 | 301(13) | 7 |
| $ZrSiO_4$ | I4$_1$/*amd* | 2.198 | 4 | 215(15) | 51, 65, 66 |
| $LaNbO_4$ | I4$_1$/*a* | 2.505 | 3 | 111(3) | 67 |
| $YVO_4$ | I4$_1$/*a* | 2.387 | 3 | 138(9) | 64 |
| $TbVO_4$ | I4$_1$/*amd* | 2.369 | 3 | 149(5) | 68 |
| $BiVO_4$ | I4$_1$/*a* | 2.350 | 3 | 150(5) | 69 |
| $DyVO_4$ | I4$_1$/*amd* | 2.354 | 3 | 160(5) | 70 |
| $YVO_4$ | I4$_1$/*amd* | 2.348 | 3 | 130(3) | 64 |
| $ErVO_4$ | I4$_1$/*amd* | 2.341 | 3 | 136(9) | 71 |
| $LuPO_4$ | I4$_1$/*amd* | 2.306 | 3 | 166(9) | 72 |
| $BaSO_4$ | P*nma* | 2.879 | 2 | 58(5) | 73, 74 |
| $BaWO_4$ | I4$_1$/*a* | 2.678 | 2 | 57(4) | 13, 75 |
| $PbWO_4$ | I4$_1$/*a* | 2.579 | 2 | 64(2) | 19 |
| $PbMoO_4$ | I4$_1$/*a* | 2.576 | 2 | 64(2) | 19 |
| $SrWO_4$ | I4$_1$/*a* | 2.557 | 2 | 63(7) | This work |
| $EuWO_4$ | I4$_1$/*a* | 2.557 | 2 | 71(6) | 75 |
| $SrMoO_4$ | I4$_1$/*a* | 2.556 | 2 | 73(5) | 76 |
| $NaY(WO_4)_2$ | I4$_1$/*a* | 2.478 | 2 | 77(8) | 77 |
| $CaMoO_4$ | I4$_1$/*a* | 2.458 | 2 | 82(7) | 19, 29 |
| $CaWO_4$ | I4$_1$/*a* | 2.457 | 2 | 75(7) | This work, 8, 9, 19, 74 |
| $SrSO_4$ | P*nma* | 2.452 | 2 | 82(5) | 16 |
| $CdMoO_4$ | I4$_1$/*a* | 2.419 | 2 | 104(2) | 19 |
| $KReO_4$ | I4$_1$/*a* | 2.791 | 1 | 18(6) | 78 |
| $TlReO_4$ | P*nma* | 2.765 | 1 | 26(4) | 79 |
| $AgReO_4$ | I4$_1$/*a* | 2.524 | 1 | 31(6) | 80 |

**Table IV:** Predicted bulk modulus for different scheelite-type and zircon-type compounds.

| $ABO_4$ compound | Space Group | mean A-O bond distance [Å] | cation formal charge | $B_0$ [GPa] |
|---|---|---|---|---|
| $HfSiO_4$ | I4$_1$/*amd* | 2.186 | 4 | 235(40) |
| $YPO_4$ | I4$_1$/*amd* | 2.337 | 3 | 145(25) |
| $YAsO_4$ | I4$_1$/*amd* | 2.383 | 3 | 135(25) |
| $EuCrO_4$ | I4$_1$/*amd* | 2.410 | 2 | 87(15) |
| $ZrGeO_4$ | I4$_1$/*a* | 2.203 | 4 | 230(40) |
| $BaMoO_4$ | I4$_1$/*a* | 2.741 | 2 | 59(12) |
| $NaReO_4$ | I4$_1$/*a* | 2.446 | 1 | 42(8) |
| $KIO_4$ | I4$_1$/*a* | 2.816 | 1 | 27(5) |

## Figure captions

**Fig. 1:** The (a) scheelite, (b) wolframite, and (c) fergusonite structures of $AWO_4$ compounds. Large circles represent the A (Ca, Sr) atoms, middle-size circles correspond to the W atoms, and the small circles are the O atoms. The unit-cell, A-O bonds and W-O bonds are also shown. As a consequence of the scheelite-to-fergusonite transition two A-O and W-O bonds are enlarged (see text and **Fig. 5**); these bonds are showed as dark lines in (c). The $AO_8$ and $WO_4$ polyhedra are also shown. By comparing (a) and (c) it can be seen the polyhedra distortion caused by the scheelite-to-fergusonite transition.

**Fig. 2:** Room-temperature ADXRD data of (a) $CaWO_4$ and (b) $SrWO_4$ at different pressures up to 18 GPa. In all diagrams the background was subtracted. To better illustrate the appearance of the (020) Bragg reflection of the fergusonite structure around $2\theta \approx 4^o$ a section of the upper trace is enlarged. In the ADXRD pattern of $CaWO_4$ collected at 11.3 GPa and of $SrWO_4$ at 10.1 GPa (which are representative of the high-pressure fergusonite structure) we also show the refined profile (symbols) and the difference between the measured data and the calculated profile (dotted line). The bars indicate the calculated positions of the reflections.

**Fig. 3:** Evolution of the (a) volume and (b)-(c) lattice parameters of $CaWO_4$ and $SrWO_4$ with pressure. Empty squares correspond to our data for the scheelite phase and empty circles and diamonds to those for the fergusonite phase. Solid squares **[8]**, solid triangles **[9]**, solid circles **[19]**, stars **[47]**, and empty hexagons **[48, 49]** are other data for the scheelite phase obtained from the literature. Empty triangles are the fergusonite data reported in Ref. **[9]**. In (a) the solid lines represent the EOS of the scheelite phase described in the text.

**Fig. 4:** Pressure dependence of the axial ratios of $CaWO_4$ and $SrWO_4$. For a description of the symbols see **Fig. 3**.

**Fig. 5:** Pressure dependence of the interatomic bond distances in the scheelite phase of (a) $CaWO_4$ and (b) $SrWO_4$. Empty squares represent the distances in the scheelite phase here reported. Solid circles **[19]**, solid squares **[27]**, and solid diamonds **[52]** represent the distances in the scheelite phase reported in the literature. Empty diamonds represent the new bond distances in the fergusonite phase after the phase transition.

**Fig. 6:** *Ab initio* simulation of XANES spectra of $CaWO_4$ and $SrWO_4$ in the three phases scheelite, fergusonite, and wolframite. The main difference between the fergusonite and the wolframite phases affects the intensity of the B resonance and the intensity and width of the white line (labeled A in the figure). There are also minor intensity changes in the C, D, and E resonances.

**Fig. 7:** Experimental XANES spectra of (a) $CaWO_4$ and (b) $SrWO_4$ measured at different pressures. The spectra collected on pressure release are marked with "d". The analysis of the spectra reveals a transition to the fergusonite phase in both compounds. At the transition we observed intensity changes in the resonances. In $CaWO_4$ B decreases a 8% and the ratio between the intensities of D and E decrease a 7%.

**Fig. 8:** Energy-volume curves (both per formula unit) calculated for (a) $CaWO_4$ and (b) $SrWO_4$. The structures shown are: scheelite (circles), fergusonite (triangles), wolframite (crosses), $LaTaO_4$ (diamonds), *Cmca* (squares), $BaWO_4$-II (dots), and $YLiF_4$-Sen (stars). The insets show differences in energy with respect to the scheelite phase in the marked areas.

**Fig. 9:** Values of the ambient-pressure bulk modulus of $ABO_4$ scheelite and scheelite-related compounds plotted against the value of the cation charge density of the $AO_8$ polyhedra. A-O distances and ambient-pressure bulk moduli were taken from different references **[7-9, 13, 16, 19, 27, 51, 65 - 80]** and are summarized in Table III. The white circle represents the bulk modulus reported by Scott **[7]** for scheelite $ZrSiO_4$. The solid line corresponds to the relation given in Eq. (2) and the dashed lines indicate its lower and higher deviations.

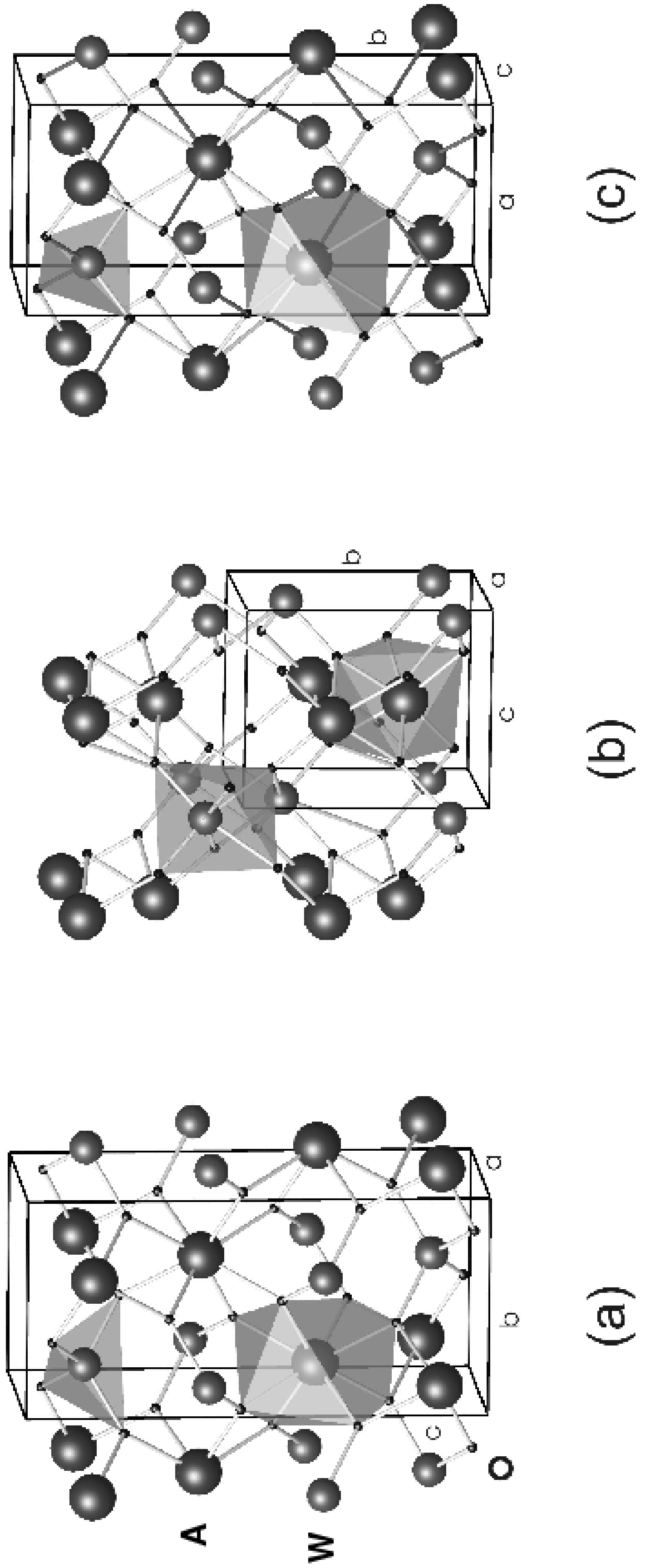

Figure 1. Errandonea *et al.*

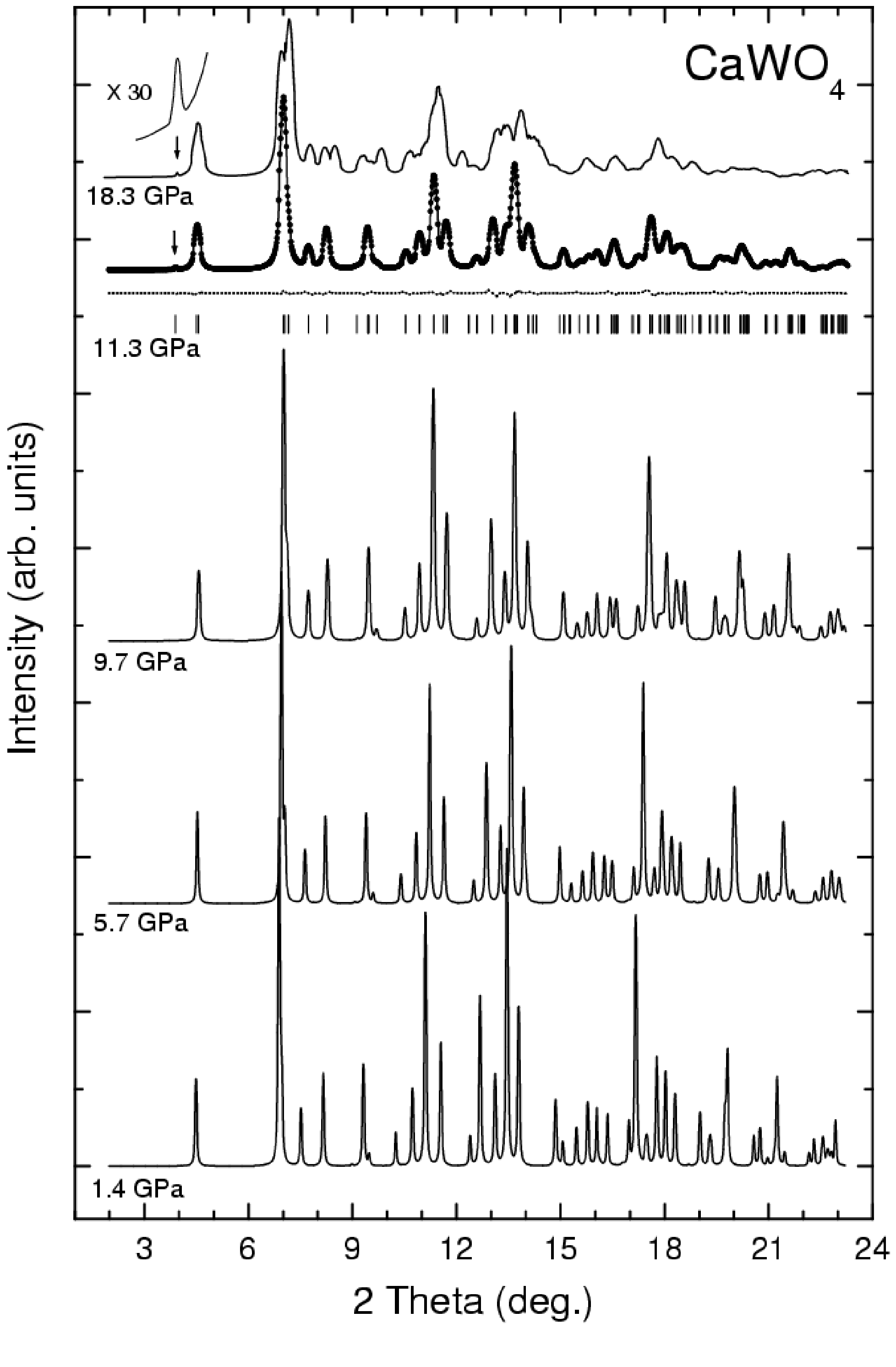

Figure 2(a) Errandonea *et al.*

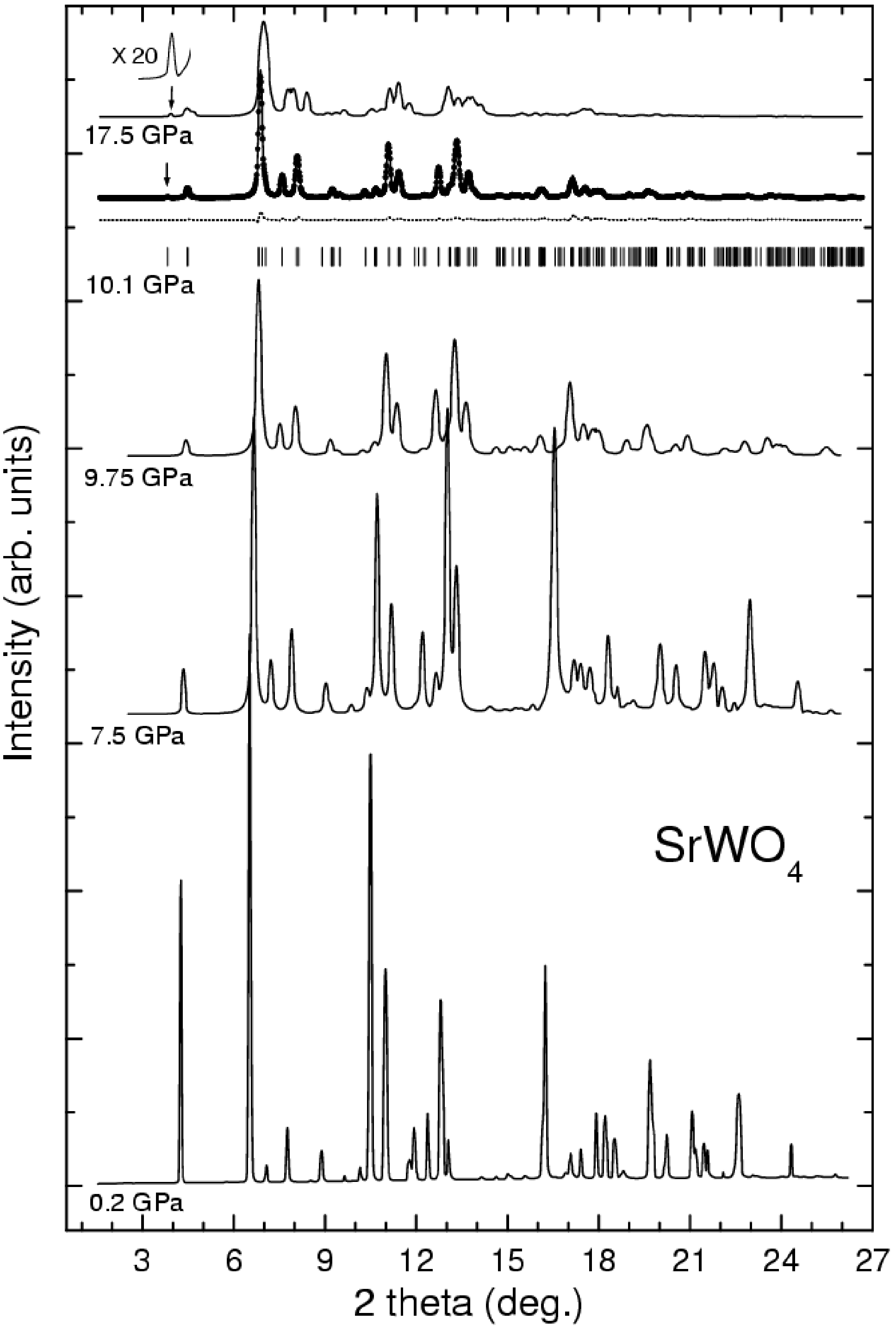

Figure 2(b) Errandonea *et al.*

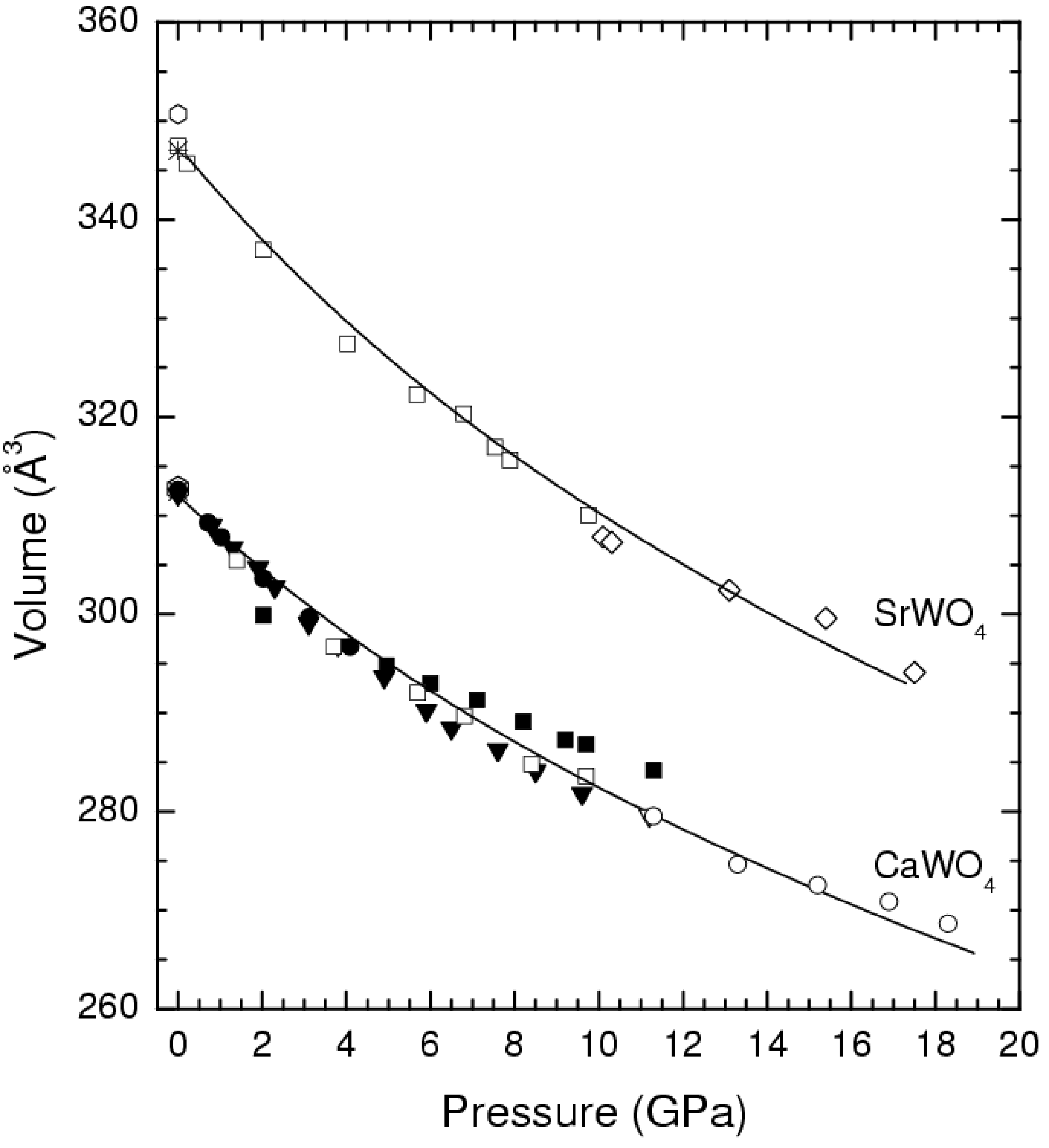

Figure 3(a) Errandonea *et al.*

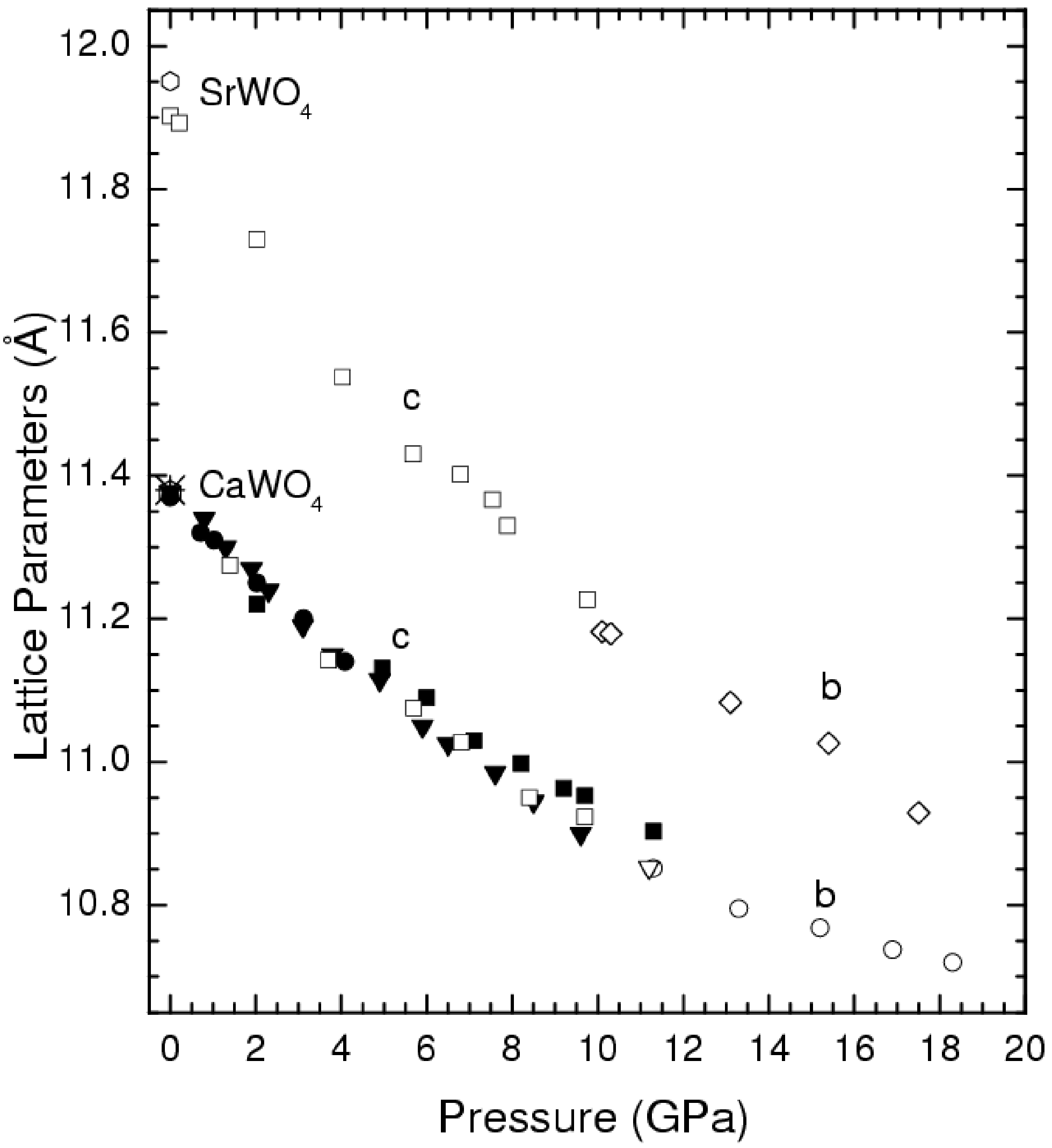

Figure 3(b) Errandonea *et al.*

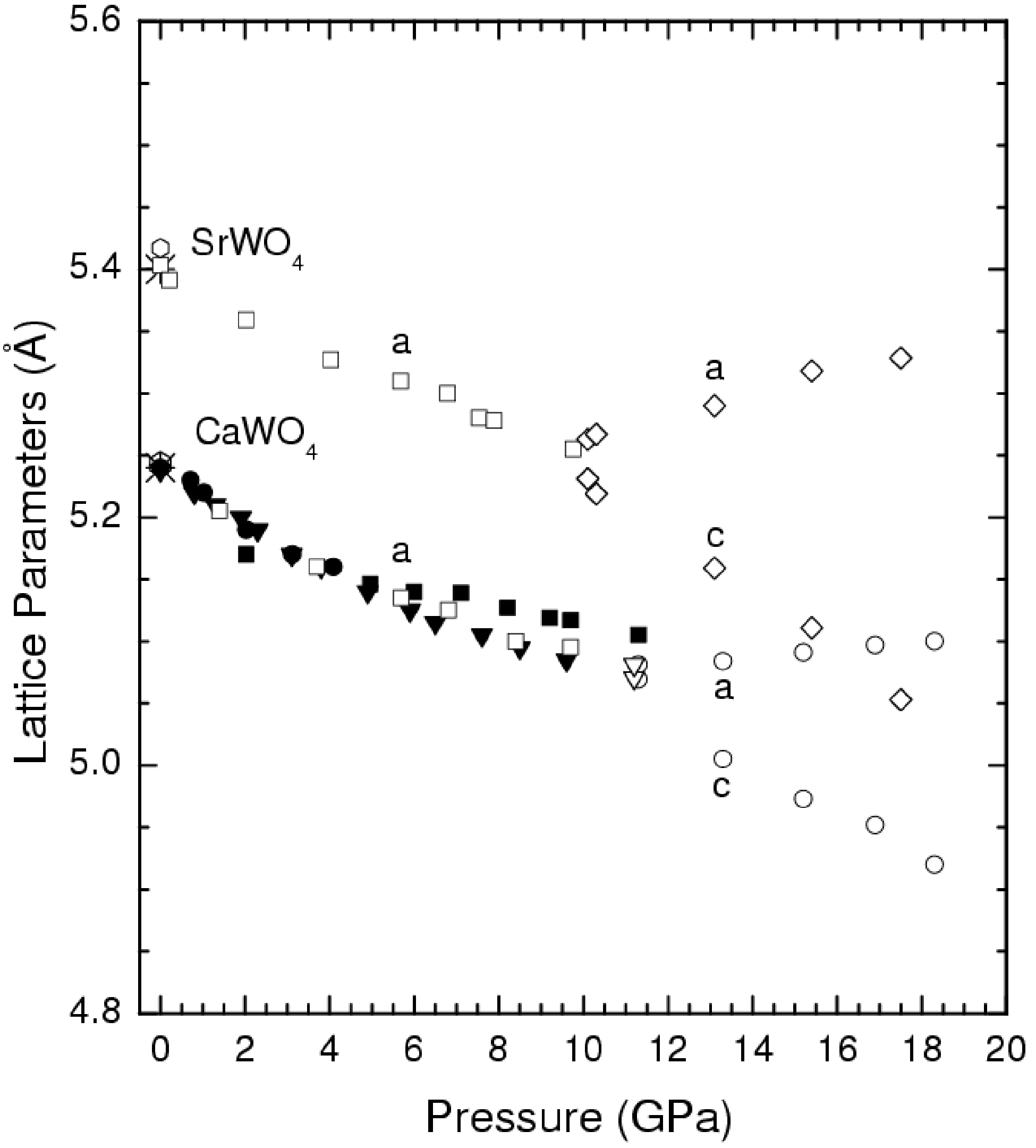

Figure 3(c) Errandonea *et al.*

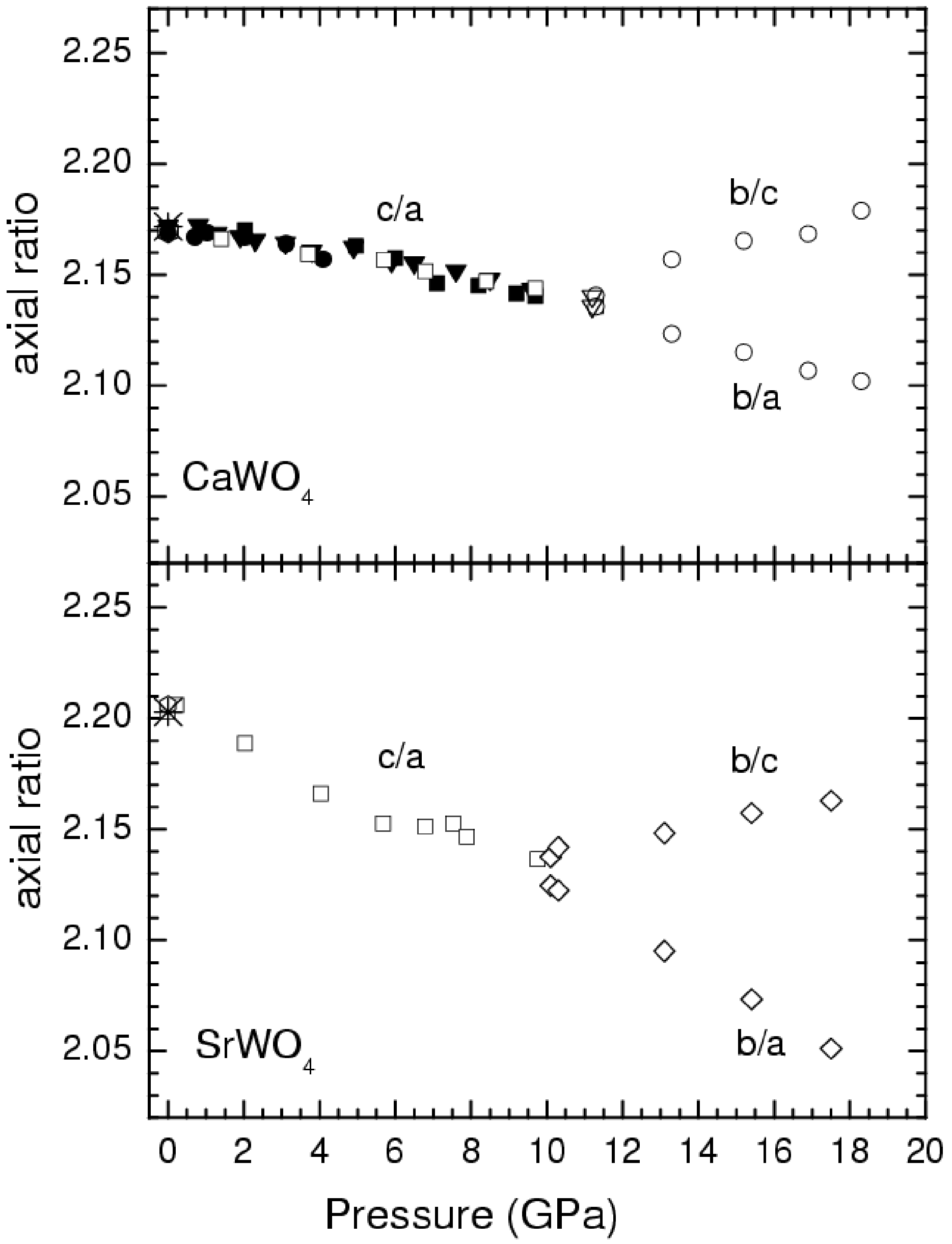

Figure 4 Errandonea *et al.*

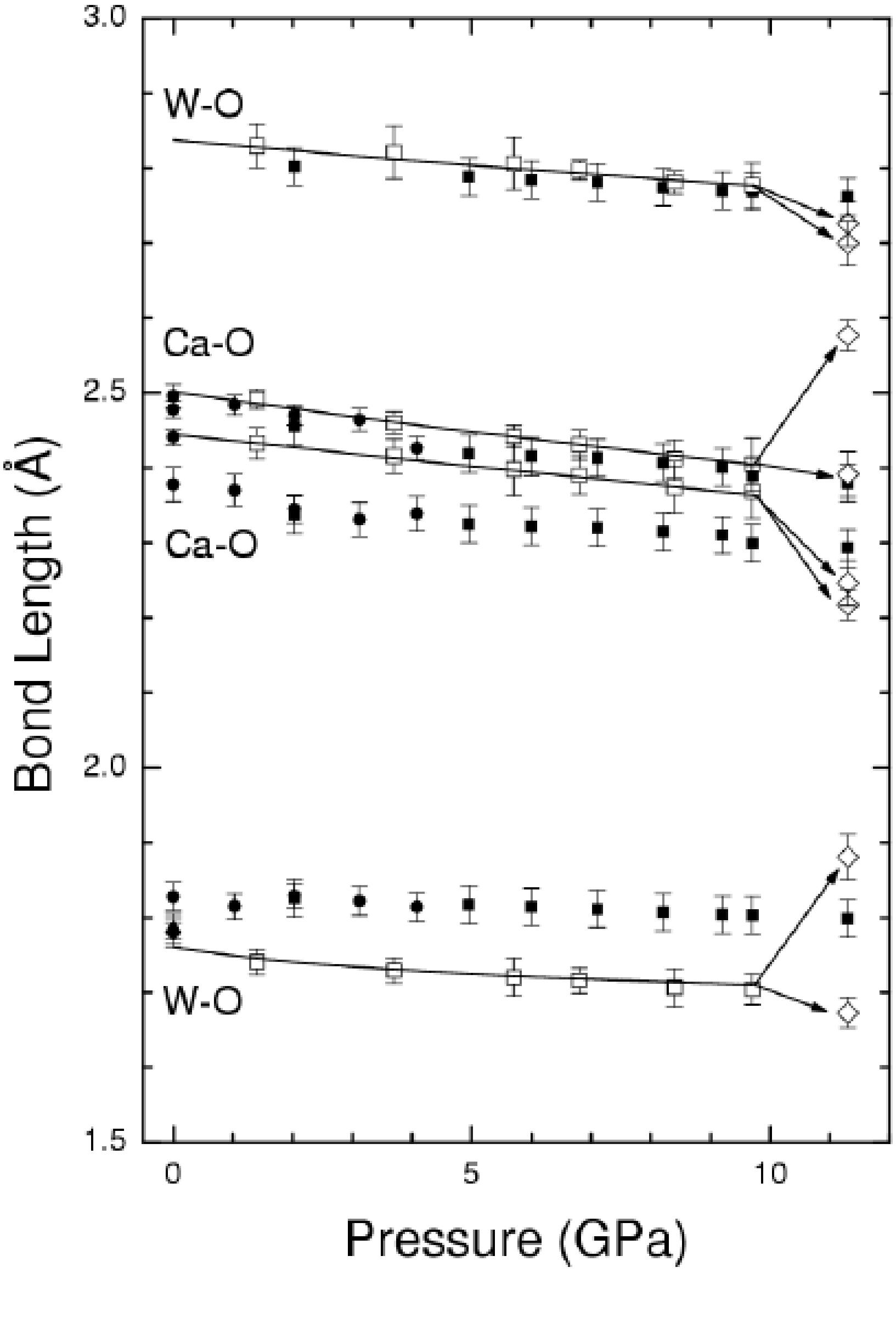

Figure 5(a) Errandonea *et al.*

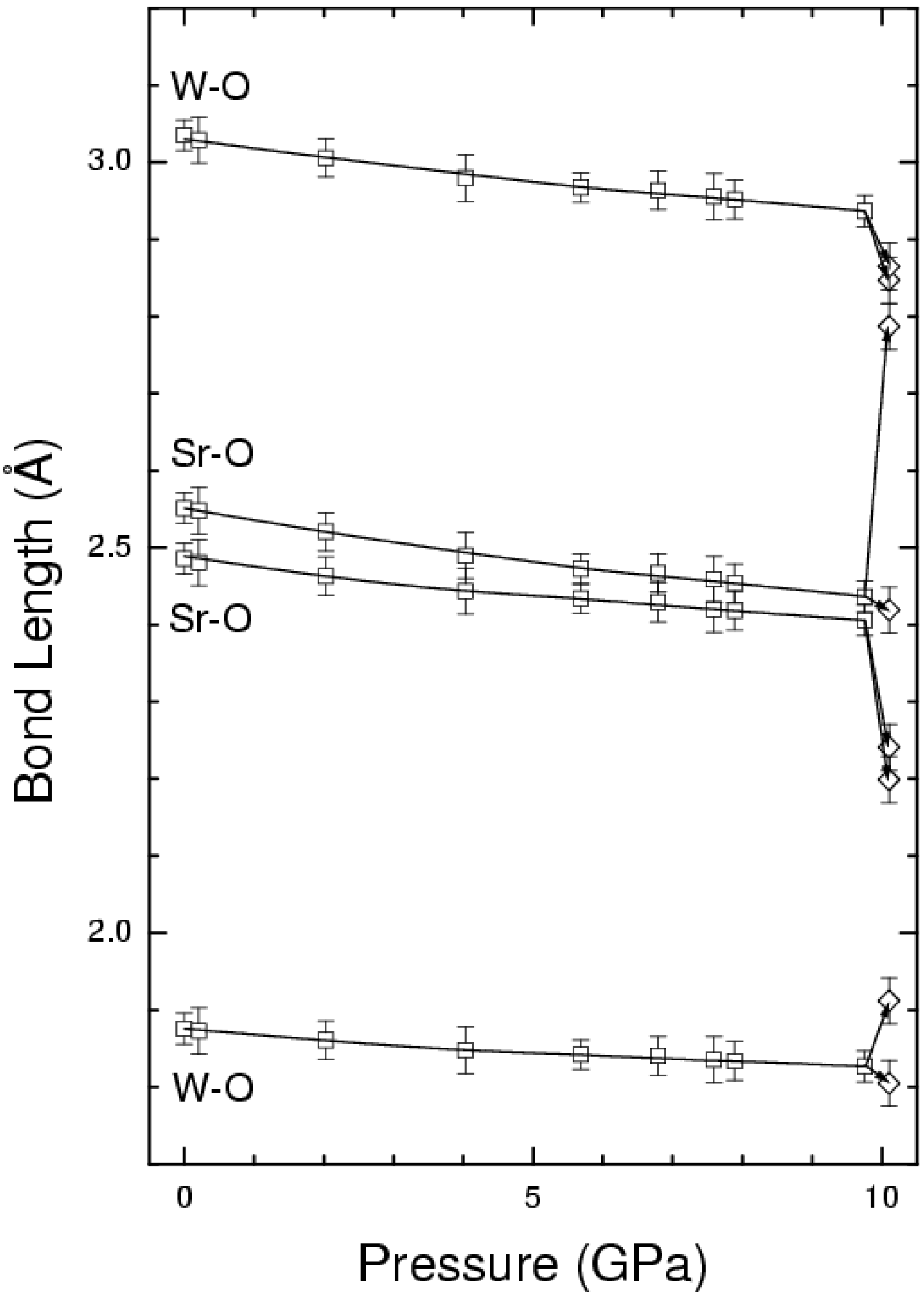

Figure 5(b) Errandonea *et al.*

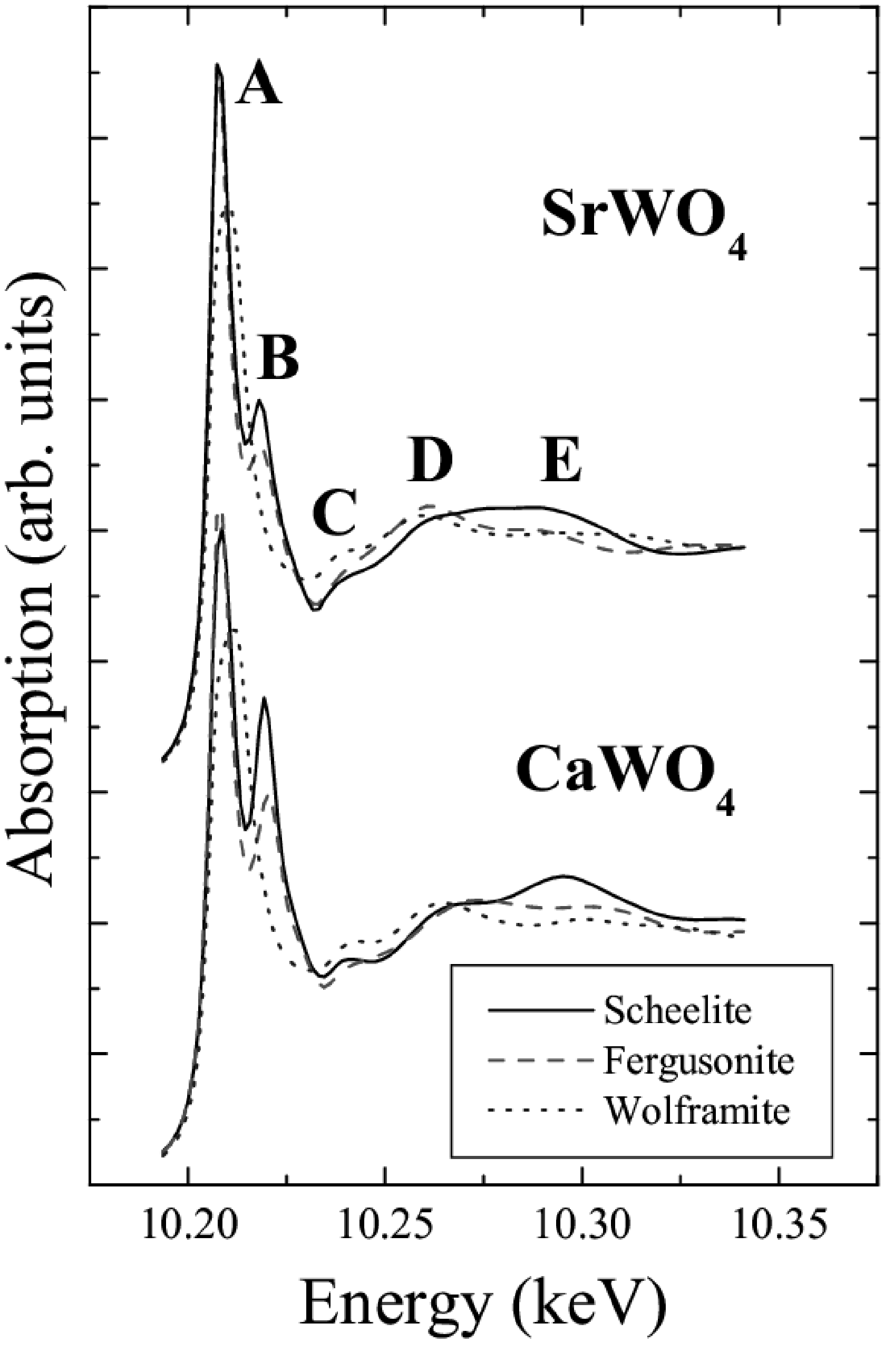

Figure 6 Errandonea *et al.*

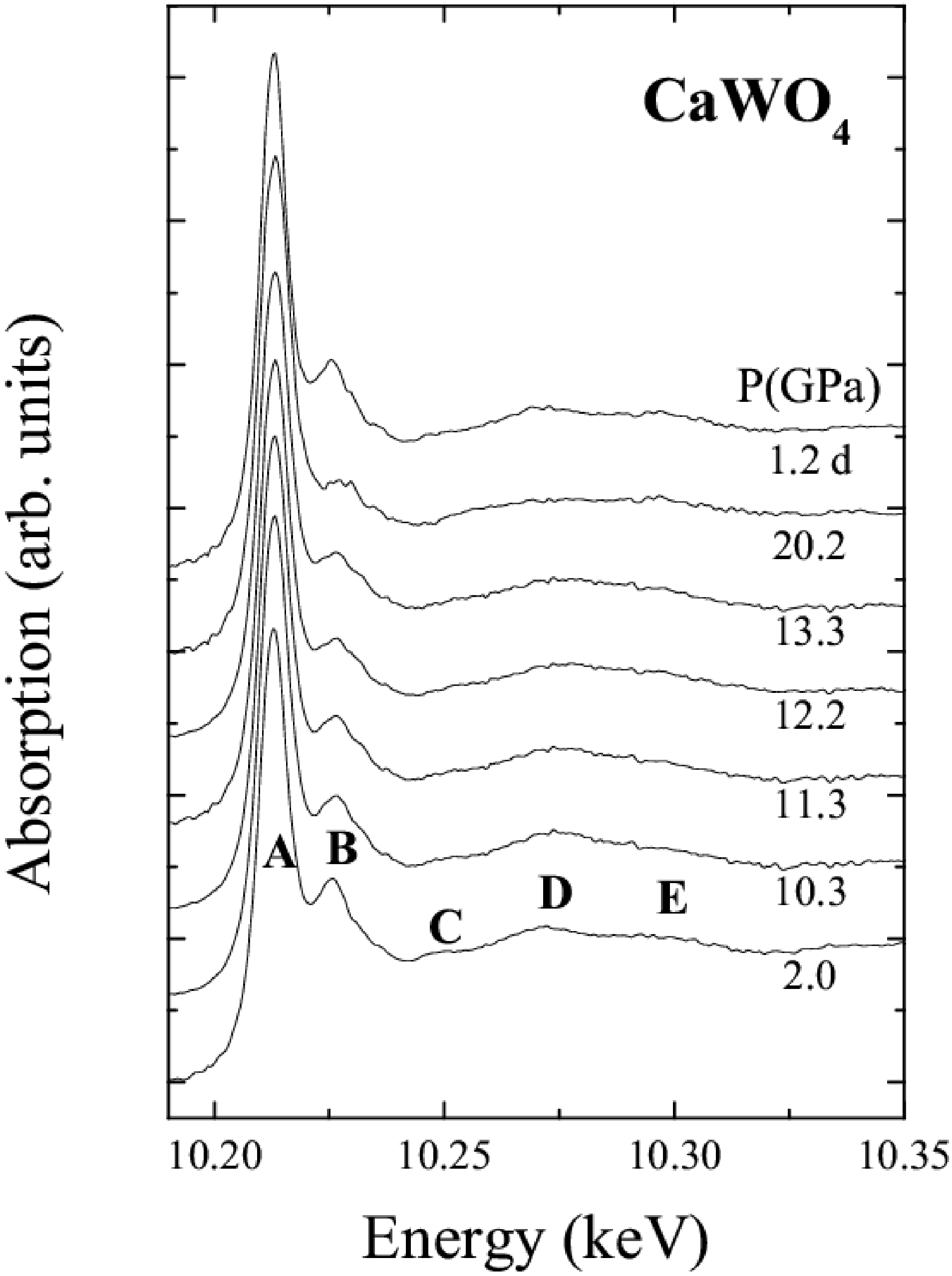

Figure 7(a) Errandonea *et al.*

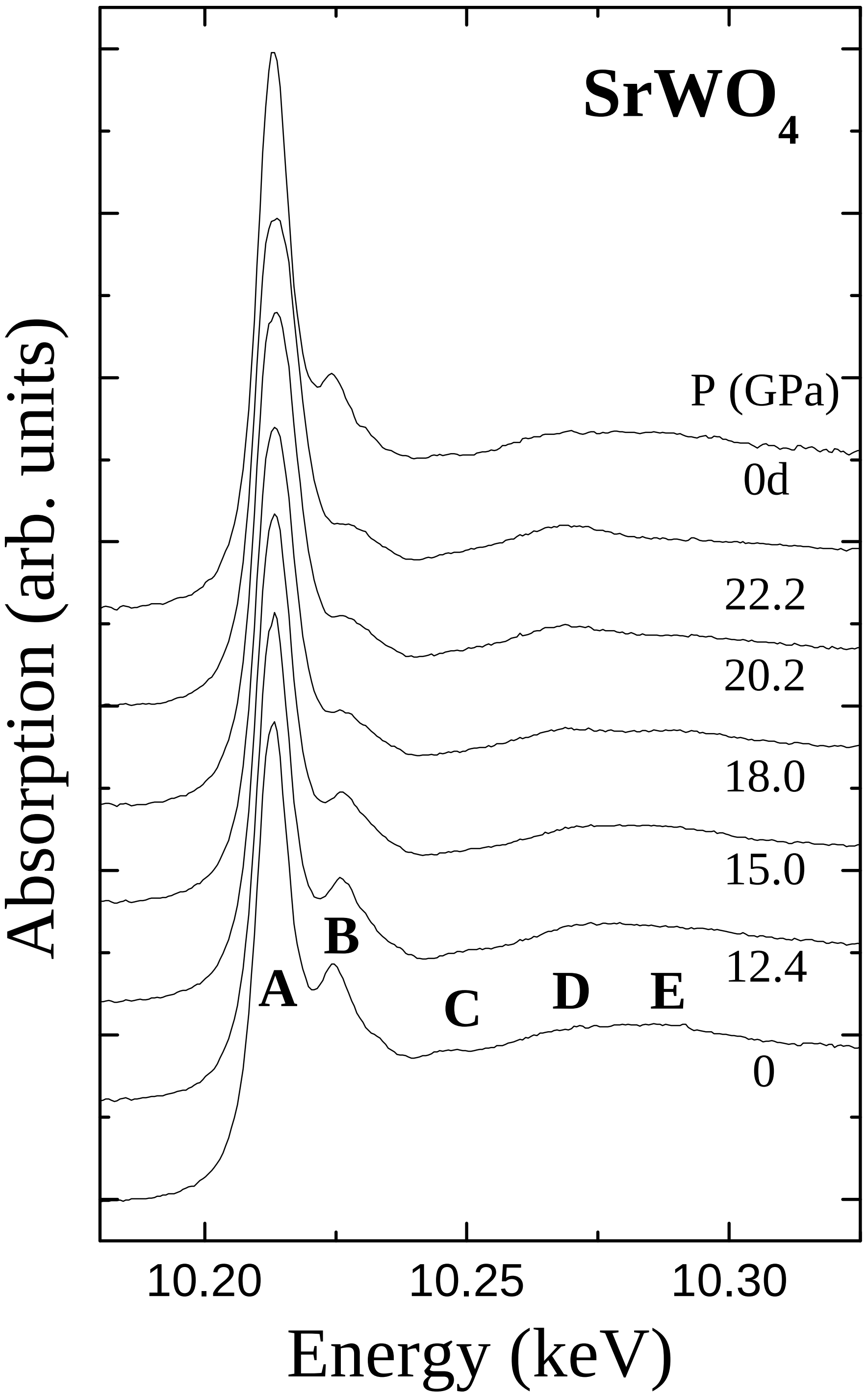

Figure 7(b) Errandonea *et al*.

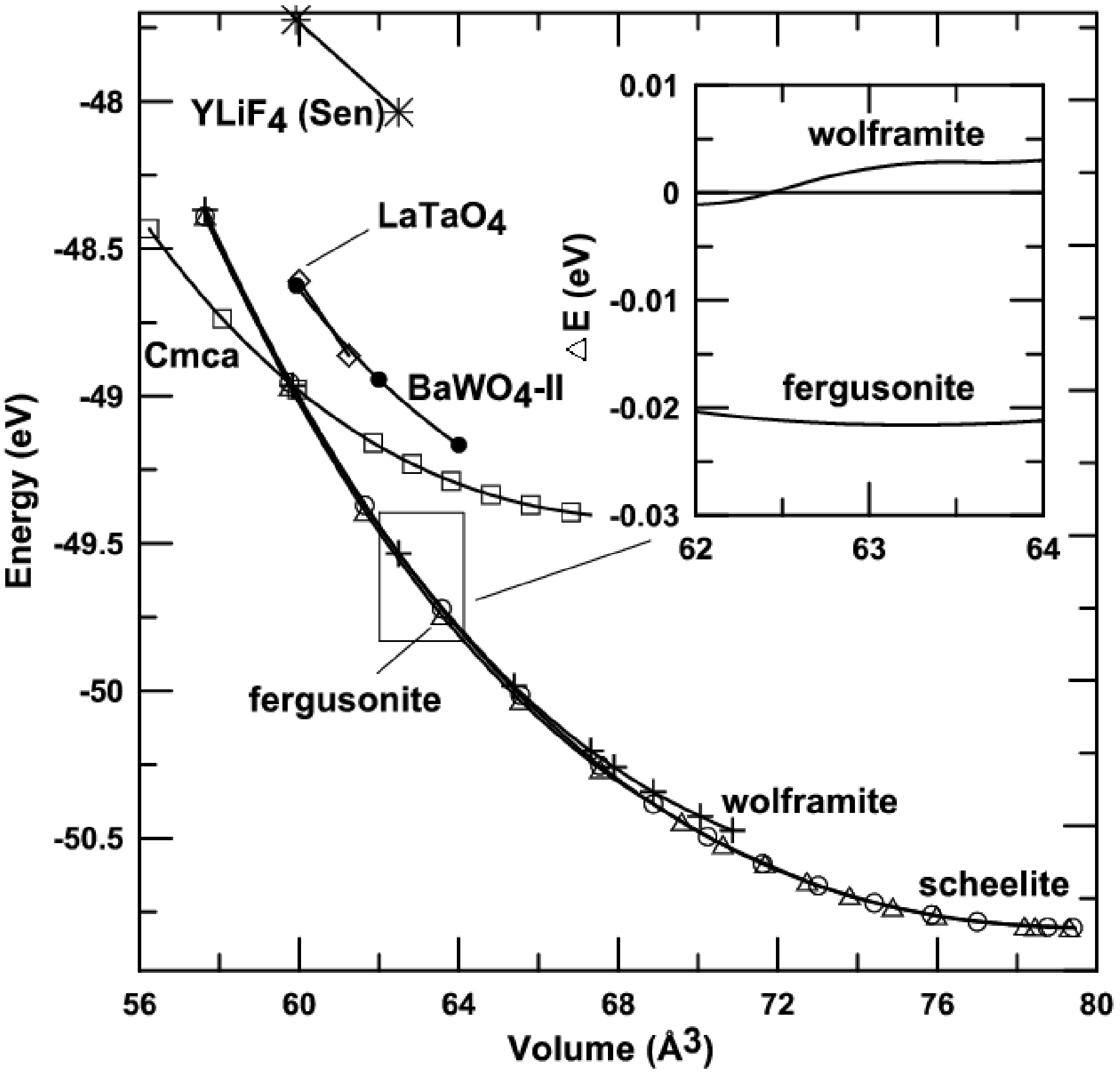

Figure 8(a) Errandonea *et al.*

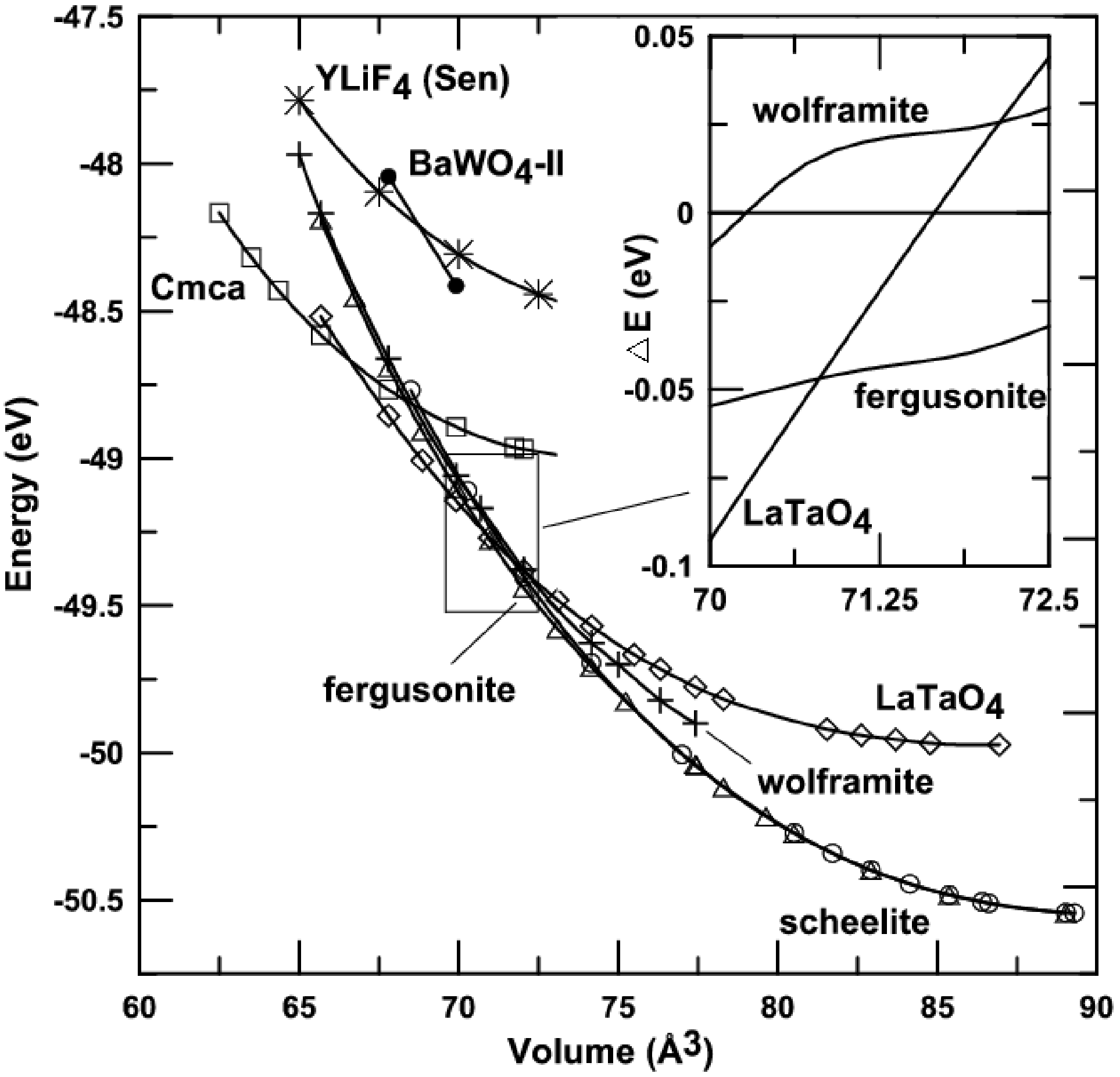

Figure 8(b) Errandonea *et al.*

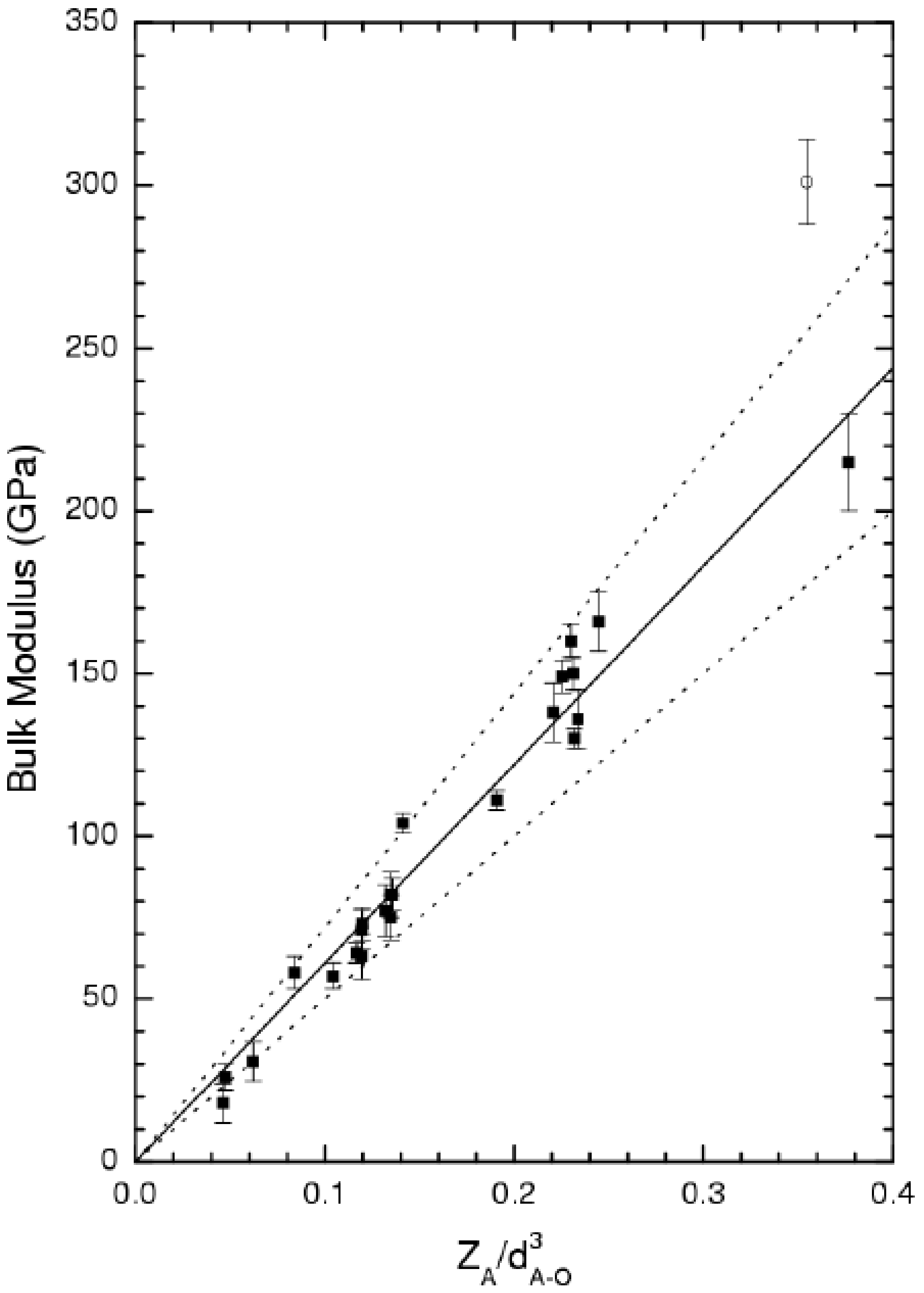

Figure 9 Errandonea *et al.*